\documentclass[%
 aip,
 amsmath,amssymb,
 reprint,%
 superscriptaddress,
 bibnotes,
 amsmath,
 amssymb,
]{revtex4-1}

\usepackage{graphicx}
\usepackage{dcolumn}%
\usepackage{bm}
\usepackage[utf8]{inputenc}
\usepackage[T1]{fontenc}
\usepackage{mathptmx}
\usepackage{etoolbox}
\usepackage{epsfig}
\usepackage{epstopdf}
\usepackage{xcolor}
\usepackage{tabularx}
\usepackage{upgreek}

\makeatletter
\def\@email#1#2{%
 \endgroup
 \patchcmd{\titleblock@produce}
  {\frontmatter@RRAPformat}
  {\frontmatter@RRAPformat{\produce@RRAP{*#1\href{mailto:#2}{#2}}}\frontmatter@RRAPformat}
  {}{}
}%
\makeatother

\begin{document}

\preprint{AIP/123-QED}

\title{Ultrarelativistic electron beams accelerated by terawatt scalable kHz laser}

\author{C.~M.~Lazzarini}
\email{CarloMaria.Lazzarini@eli-beams.eu}
\affiliation{ELI Beamlines Facility, The Extreme Light Infrastructure ERIC, Za Radnicí 835, 25241 Dolní Břežany, Czech Republic}
\affiliation{Faculty of Nuclear Sciences and Physical Engineering, Czech Technical University in Prague, Břehová 7, 11519 Prague, Czech Republic}

\author{G.~M.~Grittani}
\thanks{This author contributed equally}
\affiliation{ELI Beamlines Facility, The Extreme Light Infrastructure ERIC, Za Radnicí 835, 25241 Dolní Břežany, Czech Republic}

\author{P.~Valenta}
\affiliation{ELI Beamlines Facility, The Extreme Light Infrastructure ERIC, Za Radnicí 835, 25241 Dolní Břežany, Czech Republic}

\author{I.~Zymak}
\affiliation{ELI Beamlines Facility, The Extreme Light Infrastructure ERIC, Za Radnicí 835, 25241 Dolní Břežany, Czech Republic}

\author{R.~Antipenkov}
\affiliation{ELI Beamlines Facility, The Extreme Light Infrastructure ERIC, Za Radnicí 835, 25241 Dolní Břežany, Czech Republic}

\author{U.~Chaulagain}
\affiliation{ELI Beamlines Facility, The Extreme Light Infrastructure ERIC, Za Radnicí 835, 25241 Dolní Břežany, Czech Republic}

\author{L.~V.~N.~Goncalves }
\affiliation{ELI Beamlines Facility, The Extreme Light Infrastructure ERIC, Za Radnicí 835, 25241 Dolní Břežany, Czech Republic}

\author{A.~Grenfell}
\affiliation{ELI Beamlines Facility, The Extreme Light Infrastructure ERIC, Za Radnicí 835, 25241 Dolní Břežany, Czech Republic}

\author{M.~Lamač}
\affiliation{ELI Beamlines Facility, The Extreme Light Infrastructure ERIC, Za Radnicí 835, 25241 Dolní Břežany, Czech Republic}
\affiliation{Faculty of Mathematics and Physics, Charles University, Ke Karlovu 3, 12116 Prague, Czech Republic}
   
\author{S.~Lorenz}
\affiliation{ELI Beamlines Facility, The Extreme Light Infrastructure ERIC, Za Radnicí 835, 25241 Dolní Břežany, Czech Republic}
\affiliation{Faculty of Nuclear Sciences and Physical Engineering, Czech Technical University in Prague, Břehová 7, 11519 Prague, Czech Republic}

\author{M.~Nevrkla}
\affiliation{ELI Beamlines Facility, The Extreme Light Infrastructure ERIC, Za Radnicí 835, 25241 Dolní Břežany, Czech Republic}
\affiliation{Faculty of Nuclear Sciences and Physical Engineering, Czech Technical University in Prague, Břehová 7, 11519 Prague, Czech Republic}
   
\author{A.~Špaček}
\affiliation{ELI Beamlines Facility, The Extreme Light Infrastructure ERIC, Za Radnicí 835, 25241 Dolní Břežany, Czech Republic}
\affiliation{Faculty of Nuclear Sciences and Physical Engineering, Czech Technical University in Prague, Břehová 7, 11519 Prague, Czech Republic}
   
\author{V.~Šobr}
\affiliation{ELI Beamlines Facility, The Extreme Light Infrastructure ERIC, Za Radnicí 835, 25241 Dolní Břežany, Czech Republic}
   
\author{W.~Szuba}
\affiliation{ELI Beamlines Facility, The Extreme Light Infrastructure ERIC, Za Radnicí 835, 25241 Dolní Břežany, Czech Republic}
   
\author{P.~Bakule}
\affiliation{ELI Beamlines Facility, The Extreme Light Infrastructure ERIC, Za Radnicí 835, 25241 Dolní Břežany, Czech Republic}
   
\author{G.~Korn}
\affiliation{ELI Beamlines Facility, The Extreme Light Infrastructure ERIC, Za Radnicí 835, 25241 Dolní Břežany, Czech Republic}

\author{S.~V.~Bulanov}
\affiliation{ELI Beamlines Facility, The Extreme Light Infrastructure ERIC, Za Radnicí 835, 25241 Dolní Břežany, Czech Republic}
\affiliation{Kansai Photon Science Institute, National Institutes for Quantum Science and Technology, 8-1-7 Umemidai, Kizugawa, Kyoto 619-0215, Japan}

\date{\today}

\begin{abstract}
We show the laser-driven acceleration of unprecedented, collimated ($ 2 \ \mathrm{mrad} $ divergence), and quasi-monoenergetic ($ 25 \ \% $ energy spread) electron beams with energy up to $ 50 \ \mathrm{MeV} $ at $ 1 \ \mathrm{kHz} $ repetition rate. The laser driver is a multi-cycle ($ 15 \ \mathrm{fs} $) $ 1 \ \mathrm{kHz} $ optical parametric chirped pulse amplification (OPCPA) system, operating at $ 26 \ \mathrm{mJ} $ ($ 1.7 \ \mathrm{TW} $). The scalability of the driver laser technology and the electron beams reported in this work pave the way towards developing high-brilliance x-ray sources for medical imaging, innovative devices for brain cancer treatment, and represent a step towards the realization of a kHz GeV electron beamline.
\end{abstract}

\maketitle

Electron accelerators are a cornerstone technology of modern society. In their variations they are daily used as strategic tools at the disposal of the healthcare system for medical imaging and cancer treatment. As an example, medical linear accelerators already use electron beams with energies up to $ 20 \ \mathrm{MeV} $ for cancer radiotherapy \cite{citrin2017}. Radio-frequency acceleration is the backbone of the technology driving these machines and it is nowadays, after its first demonstration in 1928 \cite{wideroee1928}, generally acknowledged as an extremely reliable technology. Radio-frequency acceleration technology has nevertheless a fundamental limitation that is to be found in their maximum achievable accelerating gradient ($ \approx 100 \ \mathrm{MV/m} $). 

As of today, various proven technologies with the potential to overcome these limitations (by withstanding accelerating gradients exceeding $ 100 \ \mathrm{MV/mm} $ and drastically reducing the accelerator footprint) exist, such as laser-driven wakefield acceleration  (LWFA) \cite{tajima1979, esarey2009}, particle-driven (or plasma) wakefield acceleration (PWFA) \cite{chen1985, joshi2020}, structure-based wakefield acceleration \cite{oshea2016, jing2022}, or a combination of those \cite{kurz2021, foerster2022}. In recent years, the huge wakefield acceleration potential motivated an extensive effort by numerous groups worldwide, leading to remarkable achievements, ranging from the demonstration of the first GeV laser-driven electron beams \cite{leemans2006, wang2013, kim2017} to the acceleration of multi-GeV electrons via plasma guiding \cite{gonsalves2019, miao2022} and the demonstration of free electron lasing \cite{wang2021}. 

The desire to translate these exceptional results into working machines resulted in the demonstration of stable electron beam operation over more than 24 hours \cite{maier2020} and the acceleration of quasi-monoenergetic electron beams at $ 1 \ \mathrm{kHz} $ repetition rate  \cite{guenot2017, salehi2021}. With the typical charge per pulse of the order of pC \cite{goers2015, gustas2018, salehi2017,faure2018}, the resulting average current of a kHz LWFA is of the order of nA, which (provided the electron beam energy is $ > 40 \ \mathrm{MeV} $) would be enough to enable several medical applications \cite{brummer2020, svendsen2021}. These groundbreaking achievements were obtained by exploiting nonlinearities through gas-filled hollow core fiber compression down to single-cycle laser pulses in order to reach the relativistic intensity on target required to drive the wake wave in the plasma, where the injected electrons can be accelerated up to $ 15 \ \mathrm{MeV} $ \cite{salehi2021}. Despite producing very stable electron beams, this technique is limited both by the maximum laser power available and by the maximum coupling efficiency into the hollow core fiber, setting a limit to the maximum electron energy attainable. Moreover, LWFA driven by single-cycle lasers is also sensitive to carrier-envelope phase effects \cite{valenta2020, huijts2022}, posing an additional limit on the achievable electron beam energy and quality.

In this letter we show, for the first time, the production of quasi-monoenergetic ($ 25 \ \% $ energy spread), collimated ($ 2 \ \mathrm{mrad} $ divergence) electron beams with energy up to $ \mathcal{E} \approx 50 \ \mathrm{MeV} $ with multi-cycle ($ 15 \ \mathrm{fs} $) laser pulses at $ 1 \ \mathrm{kHz} $ repetition rate, overcoming the requirement of single-cycle compression and proving the energy scalability of the technology. This breakthrough was achieved by ELI Beamlines' in-house development of the L1-Allegra multi-TW $ 1 \ \mathrm{kHz} $ laser \cite{antipenkov2021}, based on the OPCPA technology.

\begin{figure*}
\centering
\includegraphics[width=5.2in]{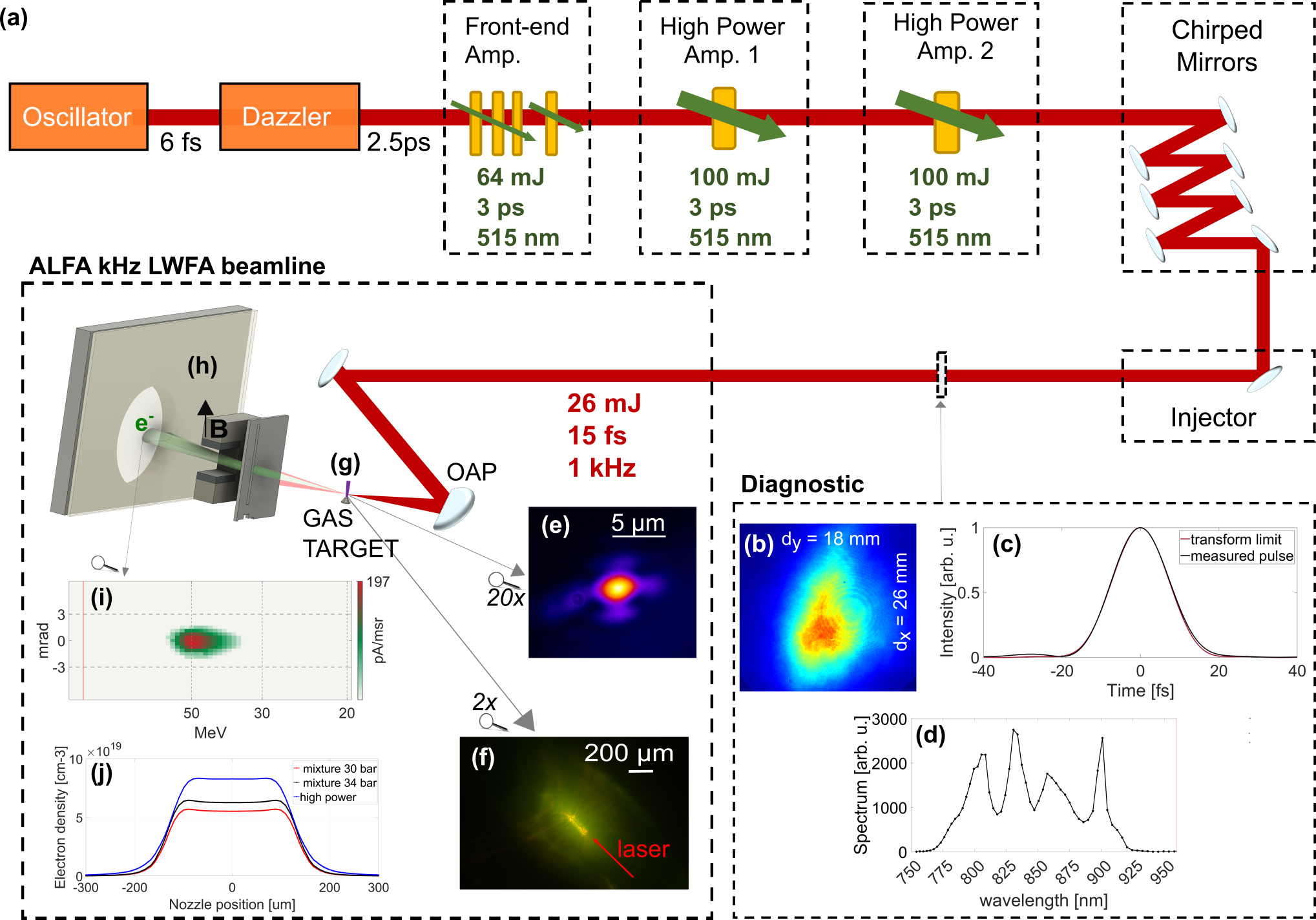}
\caption{(Color online). Simplified scheme of the experimental setup. (a) L1-Allegra laser system as used in this work, based on OPCPA stages and chirped mirror compressor. The pump sources (in green) are Yb:YAG thin-disk lasers. (b) The laser pulse near-field profile measured after the injector with Gaussian-fit diameters, giving an eccentricity $ \approx 0.8 $. (c) The pulse time duration measured by SHG FROG device compared to the transform-limited value (in red). (d)  Measured spectrum. (e) Focal spot measured with a $ 20 \times $ apochromatic microscope objective. (f) High-magnification achromatic Thomson scattering spectrally filtered (550 nm longpass) diagnostic. (g) Supersonic gas jet target. (h) Electron spectrometer. (i) Electron beam trace on the Lanex. (j) Simulated target density profiles. \label{fig:1}}
\end{figure*}

Some first demonstrations of relativistic electrons accelerated with short-pulsed OPCPA systems \cite{schmid2009, schmid2010} showed the potential of this laser technology, even if at a much lower repetition rate of $ 10 \ \mathrm{Hz} $. The L1-Allegra laser's key features for LWFA are its multi-stage power scalable design (final output expected $ > 100 \ \mathrm{mJ} $), the inherently excellent nanosecond contrast being pumped by $ 3 \ \mathrm{ps} $ Yb:YAG thin-disk lasers, and its power and pointing stability (few $ \% $ level) over many hours of continuous operation. Having a high contrast prevents unwanted detrimental effects due to pre-pulse induced plasma profile modifications.

The L1-Allegra laser was running six amplification stages [as shown in Fig.~\ref{fig:1}(a)] achieving a pulse energy of $ 26 \ \mathrm{mJ} $, measured inside the interaction chamber. Optimal compression of the output pulses was achieved through a combination of chirped mirrors and initial stretching by a programmable acousto-optic modulator (Dazzler). Ahead of the interaction chamber, the laser pulses were characterized by measuring the near-field profile [Fig.~\ref{fig:1}(b)] (where the elliptical profile is mainly due to the use of spherical mirrors in the laser telescopes and to the little amount of residual spatial chirp from the amplifiers), the time duration [Fig.~\ref{fig:1}(c)] and the spectrum [Fig.~\ref{fig:1}(d)]. These were performed by a second harmonic generation frequency-resolved optical gating (SHG FROG) device, resulting in a pulse duration of $ 15 \ \mathrm{fs} $ FWHM (with the transform-limited duration being $ 14.7 \ \mathrm{fs} $) and a central wavelength $ \lambda_{0} = 820 \ \mathrm{nm} $.

The laser pulses were focused by a $ 76.2 \ \mathrm{mm} $ focal length off-axis parabola  (nominal $ f/1 $) down to a measured focal spot [Fig.~\ref{fig:1}(e)] with FHWM of $ 4.2 $ and $ 3.1 \ \mathrm{\upmu m} $ along the horizontal and vertical axes, respectively. This corresponds to an effective beam waist $ 3.1 \pm 0.3 \ \mathrm{\upmu m} $ and to a Rayleigh range $ \approx 37 \ \mathrm{\upmu m} $. The resulting peak intensity in the focus was $ I_{0} \approx 4.8 \times 10^{18} \ \mathrm{W/cm^2} $ (corresponding to a normalized vector potential $ a_0 \approx 1.5 $). Considering the optimal electron density for this work of $ n_e \approx 5.7 \times 10^{19} \ \mathrm{cm^{-3}} $ (which corresponds to a fraction of the critical plasma density $ n_e / n_c \approx 0.034 $), the available laser power was well above the self-focusing threshold $ \approx 0.52 \ \mathrm{TW} $. This allows for self-guiding \cite{esarey2009} inside the plasma, as is visible in the Thomson scattering diagnostic [Fig.~\ref{fig:1}(f)].

The pulses were focused in the first half of the in-house designed and characterized  \cite{lorenz2019, lorenz2020} $ 300 \ \mathrm{\upmu m} $ diameter flat-top gas jet, $ 150 \ \mathrm{\upmu m} $ above the nozzle exhaust [Fig.~\ref{fig:1}(g) and (f)]. The gas target position was optimized in three dimensions with accuracy at the level of the plasma wavelength $ \lambda_{p} \approx 4.5 \ \mathrm{\upmu m} $, leading to an optimal focusing position inside the plasma profile. The repeatability and stability of the acceleration process was assured by the laser contrast higher than $ 10^{10} $ at the ps-level (measured by the Sequoia cross-correlator), below detection limit beyond several ps from the main pulse, as well as by the laser pointing stability on target of $ 3.6 \pm 0.3 \ \mathrm{\upmu rad} $ (averaged root-mean-square).

The accelerated electron beams were characterized by a calibrated electron spectrometer [Fig.~\ref{fig:1}(h)] consisting of a motorized $ 5 \ \mathrm{mm} $ collimator aluminum slit, a motorized $ 39 \ \mathrm{mm} $ long $ 0.1 \ \mathrm{T} $ permanent magnetic dipole, and a LANEX Fast Back scintillator screen captured by using a 12-bit CMOS global shutter camera [Fig.~\ref{fig:1}(i)]. The electron beam spectra were retrieved by simulating the particle tracking in the measured three-dimensional (3D) field. The measured beam energy is conservative, being the deflection measured from the furthermost slit edge.

The limited energy available out of kHz laser systems usually requires to operate targets at relatively high plasma densities in order to enable relativistic self-focusing \cite{salehi2021}. This in turn shortens $ \lambda_p $ to a few microns and ultimately results in the need of a comparable resolution in the laser-plasma interaction diagnostics \cite{lazzarini2019}. The laser-plasma interaction was first optimized at a reduced laser power of $ 0.8 \ \mathrm{TW} $, by carefully tuning the focus position inside the gas target profile monitored with a $ 5 \times $ magnification optical side-view and a $ 2 \times $ achromatic Thomson scattering spectrally filtered (550 nm longpass) top-view diagnostics, both with a resolution smaller than $ \lambda_{p} $. The gas pressure was then tuned by an electronic valve in order to achieve simultaneously relativistic self-focusing and electron injection in the plasma wave. 

The laser pulses were delivered on target at $ 1 \ \mathrm{kHz} $ repetition rate for the whole duration of the experiment, in order to reach thermal equilibrium in all the components of the system. The gas jet opening time (synchronized with the arrival of the driving pulse by a trigger) was set, depending on the regime of operation, as either pulsed or continuous. In fact, it must be noted that the radiation level achieved by a kHz LWFA machine is well above the typical values for laser facilities. The possibility of running continuous gas flow was enabled by a double differential pumping system, on target and before the compressor, which kept the laser-driven acceleration process unaffected throughout several hours of continuous operation. Typical density profiles for the specific gas jet, along with the relevant gas types and backing pressures used, were computed using ANSYS Fluent software and are depicted in Fig.~\ref{fig:1}(j).

After obtaining quasi-monoenergetic beams with $ 0.8 \ \mathrm{TW} $, we gradually ramped the laser energy up to $ 26 \ \mathrm{mJ} $ ($ 1.7 \ \mathrm{TW} $), iteratively optimizing the gas density and the focusing position inside the target density profile. In addition, the second order spectral phase was optimized using an acousto-optic programmable dispersive filter [Dazzler in Fig.~\ref{fig:1}(a)] to compensate for the plasma medium dispersion and to extend the acceleration.

\begin{figure}[t]
\includegraphics[width=0.9\linewidth]{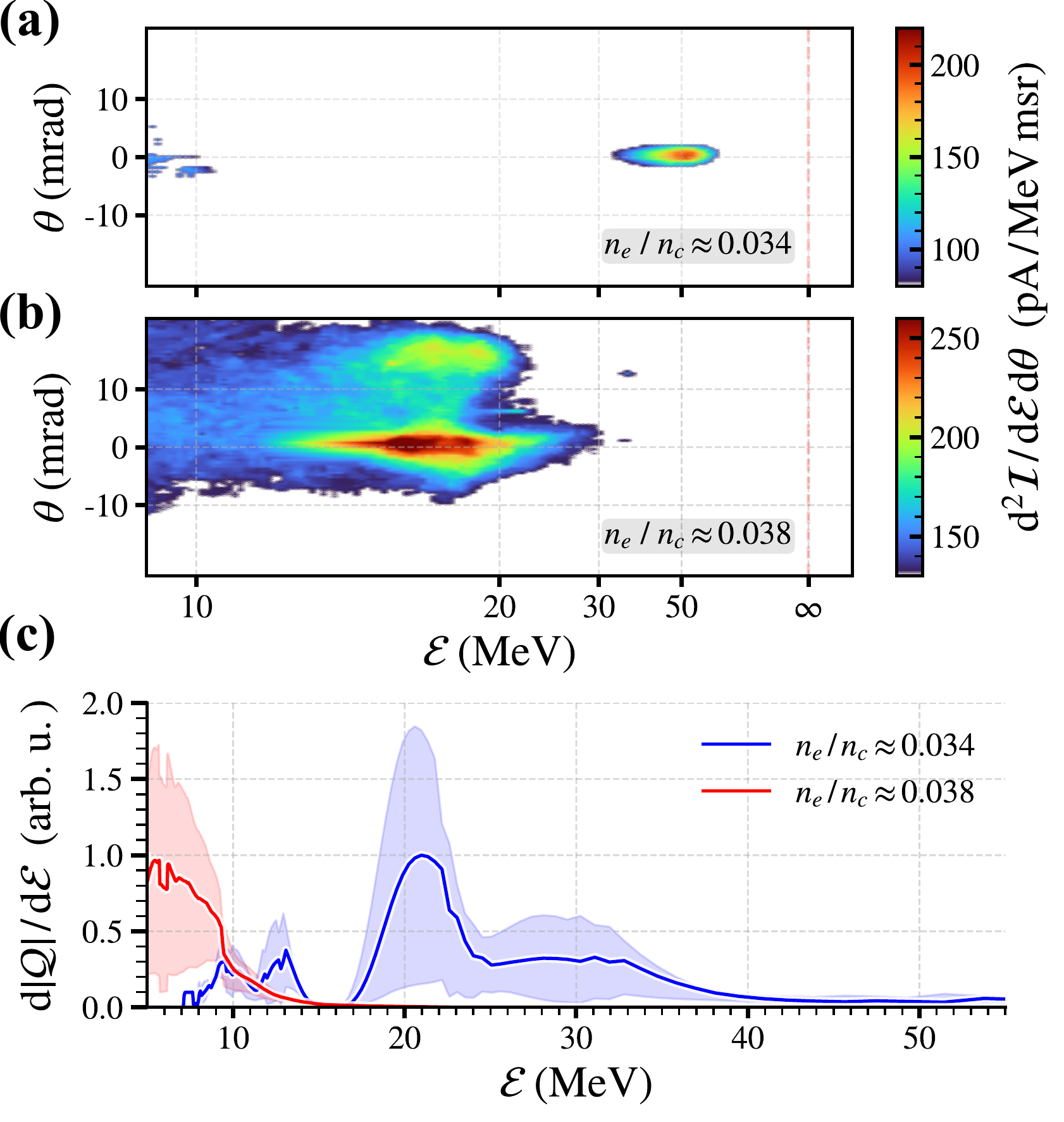}
\caption{(Color online). Transition to high-energy quasi-monoenergetic beams. (a) Selected "high-energy" spectrum, showing quasi-monoenergetic characteristics, for the nitrogen--helium mixture target operating at a plasma density $ n_e / n_c \approx 0.034 $. (b) Selected spectrum for the higher density case of $ n_e / n_c \approx 0.038 $. Each trace on the spectrometer is integrated over 10 consecutive laser shots and has the background noise filtered out, i.e., it represents the electron current, $ \mathcal{I} $, per unit of energy, $ \mathcal{E} $, and solid angle, $ \theta $. (c) 1D normalized averaged spectra for the two cases above, where the shaded area represents one standard deviation.} \label{fig:2}
\end{figure}

The most energetic and collimated electron beams were obtained by firing the laser on a gas mixture of helium ($ 98 \ \% $) and nitrogen ($ 2 \ \% $), which allowed acceleration at an electron density at the target profile plateau as low as $ n_e / n_c \approx 0.034 $. In this configuration, record-high energy (up to $ 50 \ \mathrm{MeV} $) electron beams were obtained [Fig.~\ref{fig:2}(a)]. By averaging over thousands of laser shots, we observed the high-energy quasi-monoenergetic distribution to have an average peak of $ 32 \pm 5 \ \mathrm{MeV} $, with an average energy spread of $ \approx 8 \ \mathrm{MeV} $ ($ 25 \ \% $) FWHM and a beam divergence of $ 2.1 \pm 0.8 \ \mathrm{mrad} $ FWHM.

The experimental measurements are in agreement with theoretically estimated values. Considering the ``best'' focusing scenario, where all the available laser energy is concentrated within the focal spot, the anticipated maximum electron energy is $ \approx 55 \ \mathrm{MeV} $. However, taking into account the ``average'' experimental focusing performance [Fig.~\ref{fig:1}(e)], the maximum energy drops to $ \approx 33 \ \mathrm{MeV} $.

We observed that small changes in certain laser and plasma parameters can significantly alter the LWFA process away from its optimal configuration. For example, a change in the plasma density from $ n_e / n_c \approx 0.034 $ to $ \approx 0.038 $ ($ 13 \ \% $) results in a significant reduction of the beam energy and in the loss of the quasi-monoenergetic feature [shown as example in Fig.~\ref{fig:2}(b)]. This transition is readily discernible from the average one-dimensional (1D) electron spectra depicted in Fig.~\ref{fig:2}(c) for both density profiles.

We investigated the electron acceleration for the ``average'' experimental laser and plasma parameters also with 3D particle-in-cell (PIC) simulations using the EPOCH \cite{arber2015} code. The simulations which used the mixture of a neutral gas obtained from a hydrodynamic simulation of the gas jet and incorporated the effect of field ionization shown a negligible contribution from the electrons injected via the ionization mechanism. Therefore, the laser pulse was set to propagate in a cold and collisionless plasma consisting of $ 5 \times $ ionized nitrogen and fully ionized helium [Fig.~\ref{fig:1}(j)]. The simulations were evolved over the time interval of $ 2.2 \ \mathrm{ps} $ and utilized the technique of moving window with dimensions of $ 70 \lambda_0 \times 80 \lambda_0 \times 100 \lambda_0 $. The underlying Cartesian grid was uniform with the resolution of 30 and 15 cells per $ \lambda_0 $ along the laser propagation direction and the transverse directions, respectively.

\begin{figure}[t]
\includegraphics[width=0.9\linewidth]{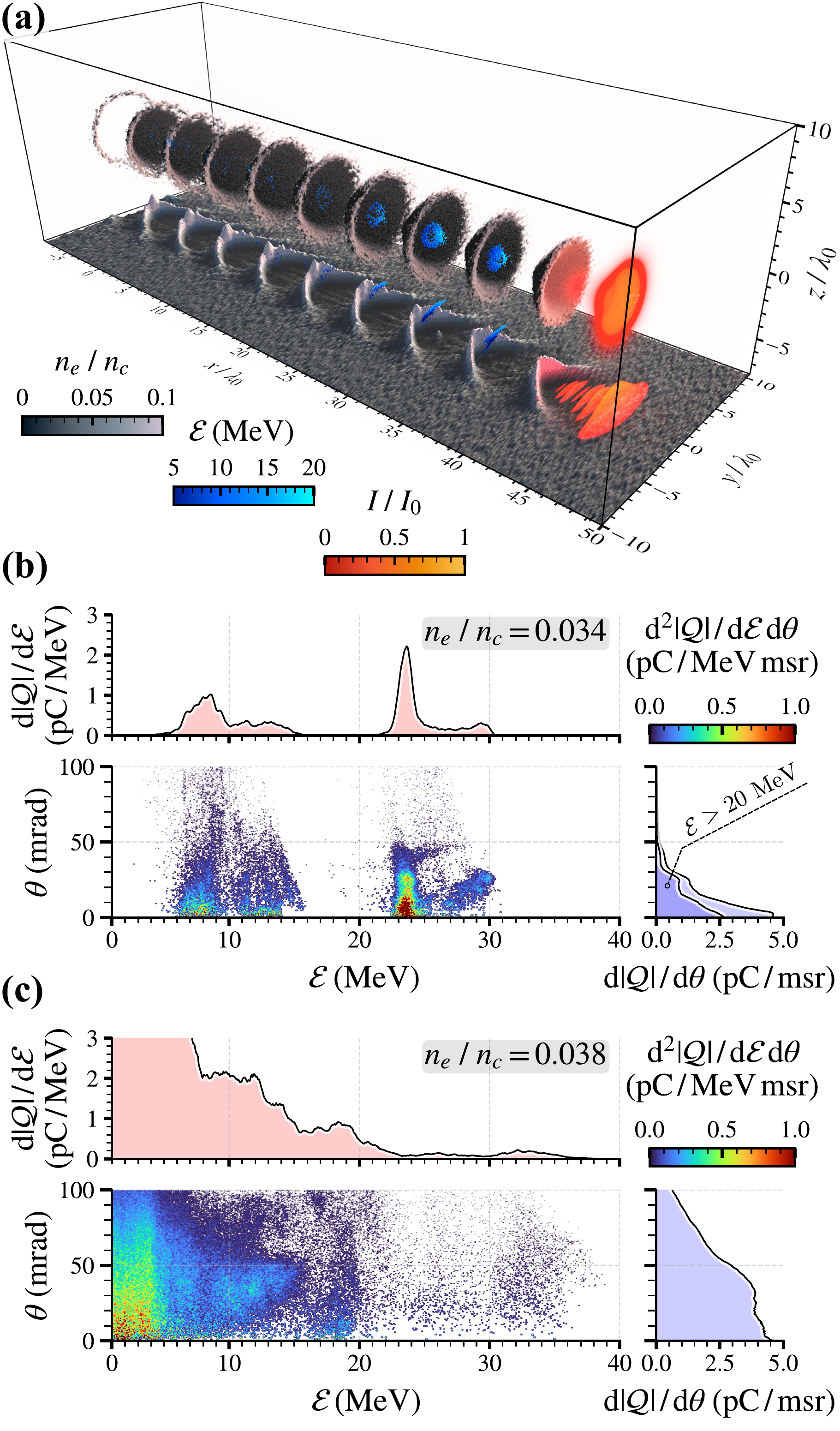}
\caption{(Color online). Results of PIC simulations. Panel (a) shows the electron number density (black to white colormap), the injected electrons with $ \mathcal{E} \geq 5 \ \mathrm{MeV} $ (dark to light blue colormap), and the laser pulse intensity (red to yellow colormap) for the $ n_e / n_c = 0.034 $ case at $ \approx 1.2 \ \mathrm{ps} $. Panels (b) and (c) show the electron energy spectra $ \mathrm{d} \left| \mathcal{Q} \right| / \mathrm{d} \mathcal{E} $ (pink), charge density of electrons with respect to the kinetic energy and propagation angle $ \mathrm{d}^2 \left| \mathcal{Q} \right| / \mathrm{d} \mathcal{E} \mathrm{d} \theta $ (blue to red colormap), and integrated charge density of electrons with respect to the propagation angle $ \mathrm{d} \left| \mathcal{Q} \right| / \mathrm{d} \theta $ including electrons with $ \mathcal{E} \geq 1 \ \mathrm{MeV} $ (light blue) and $ \mathcal{E} \geq 20 \ \mathrm{MeV} $ (dark blue), for the $ n_e / n_c = 0.034 $ and $ 0.038 $ cases, respectively, at the end of the simulations. \label{fig:3}}
\end{figure}

In the $ n_e / n_c = 0.034 $ case, the laser pulse self-focuses in the plasma and its peak amplitude attains a value $ \approx 1.7 \times $ higher than $ a_0 $. The electron self-injection occurs in the focus at several periods of the wake wave except the first one [Fig.~\ref{fig:3}(a)]. A second self-injection occurs when the laser enters the region of the electron density down-ramp, corresponding to the mechanism described in Ref.~\citenum{bulanov1998}. Consequently, two distinct electron populations can be observed in the energy spectrum [Fig.~\ref{fig:3}(b)]. The electrons with energies in the ranges $ \approx 5 - 15 $ and $ \approx 20 - 30 \ \mathrm{MeV} $ originate from the second and first injection processes, respectively. The electrons originating from the first injection process begin to dephase after $ \approx 48 \ \mathrm{\upmu m} $ of propagation within the accelerating phase of the wakefield. The wakefield structure, however, starts to slip back (with respect to the electron beam motion) due to the nonlinear evolution of the driving laser pulse such that the electrons can traverse into the accelerating phase of the preceding wakefield period. Herein, they undergo an additional energy boost, resulting in an increase of $ \approx 5 \ \mathrm{MeV} $ compared to their energy after the initial acceleration phase. At the end of the simulation, the energy spectra of both electron beams are quasi-monoenergetic. The cut-off energies of the beams are $ \approx 16.4 $ and $ \approx 31.2 \ \mathrm{MeV} $, whereas the energy spread of the higher energy beam is $ \approx 1.2 \ \mathrm{MeV} $ ($ 5 \ \% $) in FWHM. The charge and the FWHM divergence of electrons with kinetic energy $ \geq 20 \ \mathrm{MeV} $ are $ \approx 4 \ \mathrm{pC} $ and $ \approx 20 \ \mathrm{mrad} $, respectively.

By slightly reducing the electron density ($ n_e / n_c = 0.03 $), the self-injection is suppressed resulting in a negligible number of electrons being accelerated, which is in agreement with the experimental measurements. In the opposite case, corresponding to a density increase to $ n_e / n_c = 0.038 $, the laser pulse peak amplitude in the focus is $ \approx 1.8 \times $ higher than $ a_0 $. This case is characterized by fast laser energy depletion, which manifests itself in the carrier frequency downshift and slowing down of the pulse front \cite{bulanov2016}. After passing the focus, the laser splits longitudinally into two distinct pulses, each propagating at a different velocity and consisting of only a few cycles. Due to the short duration of both pulses and the relatively dense plasma, the rapid evolution of their carrier-envelope phases causes oscillations of the wake wave cavities in the laser polarization direction as well as their longitudinal modulations, strongly affecting the parameters of self-injected electrons \cite{valenta2020}. Furthermore, the electrons injected into the first period of the wake wave interact with the rear part of the laser pulse. The resulting electron energy spectrum has thermal profile [Fig.~\ref{fig:3}(c)]. The cut-off energy is $ \approx 39.2 \ \mathrm{MeV} $, the majority of electrons in the beam, however, have energy $ < 20 \ \mathrm{MeV} $. The charge of electrons with $ \mathcal{E} \geq 1 $ and $ \geq 20 \ \mathrm{MeV} $, respectively, is $ \approx 65 $ and $ 2 \ \mathrm{pC} $. The FWHM divergence of electrons with $ \mathcal{E} \geq 1 \ \mathrm{MeV} $ is $ \approx 100 \ \mathrm{mrad} $.

By further increasing the plasma density, i.e., operating at $ n_e / n_c = 0.042 $, the simulations indicate that the regime of laser-plasma interaction does not change substantially. Even though a larger fraction of electrons gets trapped, thus resulting in a higher beam charge, the pulse depletion is stronger, which reduces both the acceleration length and the maximum reachable electron energy. Overall, the PIC simulations clearly reproduce the transition from the quasi-monoenergetic to the broadband electron energy spectra with the increase of the plasma density over the same values observed experimentally (Fig.~\ref{fig:2}). 

Since the medical applications mentioned above rely on high-current electron beams at relevant energies ($ > 20 \ \mathrm{MeV} $), we optimize our LWFA source into two main modes of operation, that are freely titled ``high-energy mode'' and ``high-power mode''. The high-energy mode, described above and shown in Fig.~\ref{fig:2}(a), has the beam energy and collimation maximized by tweaking the laser-plasma interaction at the lowest possible plasma density (given the available laser power). In the high-power mode, the chosen approach is to optimize the electron beam power by working with pure nitrogen at higher plasma density of $ n_e \approx 8.4 \times 10^{19} \ \mathrm{cm^{-3}} $  (equivalent to $ n_e / n_c \approx 0.05 $), partially sacrificing the average peak energy and collimation. The fact that higher beam charge can be obtained at lower electron beam energies is likewise observed in other experiments \cite{gonsalves2019}. The key parameters of the two modes of operation are summarized in Table ~\ref{tab:1}.

With our current setup in the high-power mode, we estimate a dose rate exceeding $ 1 \ \mathrm{Gy / s} $ for beam sizes of few mm, that could lead to the first demonstration of laser-driven stereotactic radiosurgery in a similar setup to the one described in \cite{svendsen2021}, showing that approximately 600 shots were necessary to deposit $ 1 \ \mathrm{Gy} $ into $ 1 \ \mathrm{cm^3} $ target volume. While medical devices for clinical use have more stringent requirements on beam stability and variability, other applications relying solely on the dose rate and repetition rate such as space radiation hardness testing \cite{hidding2017} and radiobiology\cite{horvath2023,cavallone2021} are already possible.

\begin{table}[h]
\begin{ruledtabular} 
\resizebox{\columnwidth}{!}{
\begin{tabular}{ccc}
Feature & High-energy mode & High-power mode \\ \hline
energy & $ 32 \pm 5 \ \mathrm{MeV} $ & $ 22 \pm 2 \ \mathrm{MeV} $ \\
energy spread & $ 8 \ \mathrm{MeV} \ (25 \ \%) $ & $ 15\ \mathrm{MeV} (68 \ \%) $ \\
average current & $ 12 \pm 6 \ \mathrm{pA} $ & $ 276 \pm 28 \ \mathrm{pA} $ \\
divergence & $ 2.1 \pm 0.8 \ \mathrm{mrad} $ & $ 7.8 \pm 1.2 \ \mathrm{mrad} $ \\
pointing (std) & $ 6.9 \ \mathrm{mrad} $ & $ 4.2 \ \mathrm{mrad} $ \\
power & $ 0.4 \ \mathrm{mW} $ & $ 6 \ \mathrm{mW} $ \\
\end{tabular}}
\caption{Averaged electron beam parameters for the high-energy (corresponding to the left column of Fig.~\ref{fig:2}) and the high-power (for $ n_e / n_c \approx 0.05 $) modes. \label{tab:1}}
\end{ruledtabular}
\end{table}

In conclusion, we demonstrate the generation of quasi-monoenergetic electron beams at $ 1 \ \mathrm{kHz} $ repetition rate with an unprecedented energy of tens of MeV (up to $ 50 \ \mathrm{MeV} $) which paves the way towards establishing LWFA technology as an innovative tool for treatment of small tumors (e.g., brain metastasis) and for the generation of high-flux monoenergetic x-ray beams in the medical imaging range ($ 20 - 50 \ \mathrm{keV} $). The latter may represent a unique source for x-ray fluorescent imaging (XFI), phase contrast imaging (PCI), and micro-beam radiotherapy. Consistent work is ongoing to produce a stable and reliable high-energy ($ > 100 \ \mathrm{MeV} $), high-flux electron beamline for innovative user experiments. Finally, we want to highlight that the all-reflection LWFA setup based on OPCPA technology, shown in this work, is fully scalable in laser power. This paves the way towards the realization of a future GeV-class $ \mathrm{kHz} $ electron beamline.

\begin{acknowledgments}
We thank H. Milchberg for fruitful discussions. We acknowledge helpful feedback on the manuscript from E. Chacon Golcher and J. Limpouch. We also acknowledge M. Favetta, G. Tasselli, F.I.M. Fucilli, and M. Piombino for the availability of the medical linear accelerator used to calibrate the electron spectrometer.

This work was supported by the project ``ADONIS - Advanced research using high-intensity photons and particles'' (CZ.02.1.01/0.0/0.0/16\_019/0000789) from the European Regional Development Fund, by the ``IMPULSE'' project, which receives funding from the European Union Framework Programme for Research and Innovation Horizon 2020 under Grant Agreement No. 871161, and by the project ``e-INFRA CZ'' (ID:90254) from the Ministry of Education, Youth and Sports of the Czech Republic. The EPOCH code used in this work was in part funded by the UK EPSRC grants EP/G054950/1, EP/G056803/1, EP/G055165/1, EP/M022463/1, and EP/P02212X/1.
\end{acknowledgments}

\section*{AUTHOR DECLARATIONS}

\subsection*{Conflict of Interest} 
The authors have no conflicts to disclose.

\subsection*{Author Contributions}
R.A., P.B., A.G., V.S., A.S., and W.S. are the L1-Allegra laser team. G.K. had the idea for this experiment. The ALFA LWFA beamline was conceived and designed by C.M.L. and G.M.G. and its testing and full integration to the laser system was performed by C.M.L., G.M.G., I.Z., S.L., M.N., L.V.N.G., R.A., P.B, and V.S. The optical layout of the accelerator was designed and installed by C.M.L. with the help of M.N., L.V.N.G., G.M.G., and I.Z. Theoretical analysis and simulations were performed by P.V. and S.V.B. The experiment was coordinated by C.M.L. and G.M.G. All the data analysis was performed by I.Z., C.M.L., and G.M.G. Written and prepared by C.M.L., G.M.G., P.V., I.Z., A.S., G.K., and S.V.B.

\section*{DATA AVAILABILITY}
The data that support the findings of this study are available from the corresponding author upon reasonable request. \\

This article may be downloaded for personal use only. Any other use requires prior permission of the author and AIP Publishing. This article appeared in C.M. Lazzarini et al., Phys. Plasmas 31, 030703 (2024) and may be found at \url{https://doi.org/10.1063/5.0189051}.

\nocite{*}

\section*{REFERENCES}

\begin{thebibliography}{40}%
\makeatletter
\providecommand \@ifxundefined [1]{%
 \@ifx{#1\undefined}
}%
\providecommand \@ifnum [1]{%
 \ifnum #1\expandafter \@firstoftwo
 \else \expandafter \@secondoftwo
 \fi
}%
\providecommand \@ifx [1]{%
 \ifx #1\expandafter \@firstoftwo
 \else \expandafter \@secondoftwo
 \fi
}%
\providecommand \natexlab [1]{#1}%
\providecommand \enquote  [1]{``#1''}%
\providecommand \bibnamefont  [1]{#1}%
\providecommand \bibfnamefont [1]{#1}%
\providecommand \citenamefont [1]{#1}%
\providecommand \href@noop [0]{\@secondoftwo}%
\providecommand \href [0]{\begingroup \@sanitize@url \@href}%
\providecommand \@href[1]{\@@startlink{#1}\@@href}%
\providecommand \@@href[1]{\endgroup#1\@@endlink}%
\providecommand \@sanitize@url [0]{\catcode `\\12\catcode `\$12\catcode `\&12\catcode `\#12\catcode `\^12\catcode `\_12\catcode `\%12\relax}%
\providecommand \@@startlink[1]{}%
\providecommand \@@endlink[0]{}%
\providecommand \url  [0]{\begingroup\@sanitize@url \@url }%
\providecommand \@url [1]{\endgroup\@href {#1}{\urlprefix }}%
\providecommand \urlprefix  [0]{URL }%
\providecommand \Eprint [0]{\href }%
\providecommand \doibase [0]{http://dx.doi.org/}%
\providecommand \selectlanguage [0]{\@gobble}%
\providecommand \bibinfo  [0]{\@secondoftwo}%
\providecommand \bibfield  [0]{\@secondoftwo}%
\providecommand \translation [1]{[#1]}%
\providecommand \BibitemOpen [0]{}%
\providecommand \bibitemStop [0]{}%
\providecommand \bibitemNoStop [0]{.\EOS\space}%
\providecommand \EOS [0]{\spacefactor3000\relax}%
\providecommand \BibitemShut  [1]{\csname bibitem#1\endcsname}%
\let\auto@bib@innerbib\@empty
\bibitem [{\citenamefont {Citrin}(2017)}]{citrin2017}%
  \BibitemOpen
  \bibfield  {author} {\bibinfo {author} {\bibfnamefont {D.~E.}\ \bibnamefont {Citrin}},\ }\bibfield  {title} {\enquote {\bibinfo {title} {Recent {{Developments}} in {{Radiotherapy}}},}\ }\href {\doibase 10.1056/NEJMra1608986} {\bibfield  {journal} {\bibinfo  {journal} {New England Journal of Medicine}\ }\textbf {\bibinfo {volume} {377}},\ \bibinfo {pages} {1065--1075} (\bibinfo {year} {2017})}\BibitemShut {NoStop}%
\bibitem [{\citenamefont {Wider{\"o}e}(1928)}]{wideroee1928}%
  \BibitemOpen
  \bibfield  {author} {\bibinfo {author} {\bibfnamefont {R.}~\bibnamefont {Wider{\"o}e}},\ }\bibfield  {title} {\enquote {\bibinfo {title} {{\"Uber ein neues Prinzip zur Herstellung hoher Spannungen}},}\ }\href {\doibase 10.1007/BF01656341} {\bibfield  {journal} {\bibinfo  {journal} {Archiv f\"ur Elektrotechnik}\ }\textbf {\bibinfo {volume} {21}},\ \bibinfo {pages} {387--406} (\bibinfo {year} {1928})}\BibitemShut {NoStop}%
\bibitem [{\citenamefont {Tajima}\ and\ \citenamefont {Dawson}(1979)}]{tajima1979}%
  \BibitemOpen
  \bibfield  {author} {\bibinfo {author} {\bibfnamefont {T.}~\bibnamefont {Tajima}}\ and\ \bibinfo {author} {\bibfnamefont {J.~M.}\ \bibnamefont {Dawson}},\ }\bibfield  {title} {\enquote {\bibinfo {title} {Laser {{Electron Accelerator}}},}\ }\href {\doibase 10.1103/PhysRevLett.43.267} {\bibfield  {journal} {\bibinfo  {journal} {Physical Review Letters}\ }\textbf {\bibinfo {volume} {43}},\ \bibinfo {pages} {267--270} (\bibinfo {year} {1979})}\BibitemShut {NoStop}%
\bibitem [{\citenamefont {Esarey}, \citenamefont {Schroeder},\ and\ \citenamefont {Leemans}(2009)}]{esarey2009}%
  \BibitemOpen
  \bibfield  {author} {\bibinfo {author} {\bibfnamefont {E.}~\bibnamefont {Esarey}}, \bibinfo {author} {\bibfnamefont {C.~B.}\ \bibnamefont {Schroeder}}, \ and\ \bibinfo {author} {\bibfnamefont {W.~P.}\ \bibnamefont {Leemans}},\ }\bibfield  {title} {\enquote {\bibinfo {title} {Physics of laser-driven plasma-based electron accelerators},}\ }\href {\doibase 10.1103/RevModPhys.81.1229} {\bibfield  {journal} {\bibinfo  {journal} {Reviews of Modern Physics}\ }\textbf {\bibinfo {volume} {81}},\ \bibinfo {pages} {1229--1285} (\bibinfo {year} {2009})}\BibitemShut {NoStop}%
\bibitem [{\citenamefont {Chen}\ \emph {et~al.}(1985)\citenamefont {Chen}, \citenamefont {Dawson}, \citenamefont {Huff},\ and\ \citenamefont {Katsouleas}}]{chen1985}%
  \BibitemOpen
  \bibfield  {author} {\bibinfo {author} {\bibfnamefont {P.}~\bibnamefont {Chen}}, \bibinfo {author} {\bibfnamefont {J.~M.}\ \bibnamefont {Dawson}}, \bibinfo {author} {\bibfnamefont {R.~W.}\ \bibnamefont {Huff}}, \ and\ \bibinfo {author} {\bibfnamefont {T.}~\bibnamefont {Katsouleas}},\ }\bibfield  {title} {\enquote {\bibinfo {title} {Acceleration of {{Electrons}} by the {{Interaction}} of a {{Bunched Electron Beam}} with a {{Plasma}}},}\ }\href {\doibase 10.1103/PhysRevLett.54.693} {\bibfield  {journal} {\bibinfo  {journal} {Physical Review Letters}\ }\textbf {\bibinfo {volume} {54}},\ \bibinfo {pages} {693--696} (\bibinfo {year} {1985})}\BibitemShut {NoStop}%
\bibitem [{\citenamefont {Joshi}, \citenamefont {Corde},\ and\ \citenamefont {Mori}(2020)}]{joshi2020}%
  \BibitemOpen
  \bibfield  {author} {\bibinfo {author} {\bibfnamefont {C.}~\bibnamefont {Joshi}}, \bibinfo {author} {\bibfnamefont {S.}~\bibnamefont {Corde}}, \ and\ \bibinfo {author} {\bibfnamefont {W.~B.}\ \bibnamefont {Mori}},\ }\bibfield  {title} {\enquote {\bibinfo {title} {Perspectives on the generation of electron beams from plasma-based accelerators and their near and long term applications},}\ }\href {\doibase 10.1063/5.0004039} {\bibfield  {journal} {\bibinfo  {journal} {Physics of Plasmas}\ }\textbf {\bibinfo {volume} {27}},\ \bibinfo {pages} {070602} (\bibinfo {year} {2020})}\BibitemShut {NoStop}%
\bibitem [{\citenamefont {O'Shea}\ \emph {et~al.}(2016)\citenamefont {O'Shea}, \citenamefont {Andonian}, \citenamefont {Barber}, \citenamefont {Fitzmorris}, \citenamefont {Hakimi}, \citenamefont {Harrison}, \citenamefont {Hoang}, \citenamefont {Hogan}, \citenamefont {Naranjo}, \citenamefont {Williams}, \citenamefont {Yakimenko},\ and\ \citenamefont {Rosenzweig}}]{oshea2016}%
  \BibitemOpen
  \bibfield  {author} {\bibinfo {author} {\bibfnamefont {B.~D.}\ \bibnamefont {O'Shea}}, \bibinfo {author} {\bibfnamefont {G.}~\bibnamefont {Andonian}}, \bibinfo {author} {\bibfnamefont {S.~K.}\ \bibnamefont {Barber}}, \bibinfo {author} {\bibfnamefont {K.~L.}\ \bibnamefont {Fitzmorris}}, \bibinfo {author} {\bibfnamefont {S.}~\bibnamefont {Hakimi}}, \bibinfo {author} {\bibfnamefont {J.}~\bibnamefont {Harrison}}, \bibinfo {author} {\bibfnamefont {P.~D.}\ \bibnamefont {Hoang}}, \bibinfo {author} {\bibfnamefont {M.~J.}\ \bibnamefont {Hogan}}, \bibinfo {author} {\bibfnamefont {B.}~\bibnamefont {Naranjo}}, \bibinfo {author} {\bibfnamefont {O.~B.}\ \bibnamefont {Williams}}, \bibinfo {author} {\bibfnamefont {V.}~\bibnamefont {Yakimenko}}, \ and\ \bibinfo {author} {\bibfnamefont {J.~B.}\ \bibnamefont {Rosenzweig}},\ }\bibfield  {title} {\enquote {\bibinfo {title} {Observation of acceleration and deceleration in gigaelectron-volt-per-metre gradient dielectric wakefield accelerators},}\ }\href {\doibase
  10.1038/ncomms12763} {\bibfield  {journal} {\bibinfo  {journal} {Nature Communications}\ }\textbf {\bibinfo {volume} {7}},\ \bibinfo {pages} {12763} (\bibinfo {year} {2016})}\BibitemShut {NoStop}%
\bibitem [{\citenamefont {Jing}\ and\ \citenamefont {Ha}(2022)}]{jing2022}%
  \BibitemOpen
  \bibfield  {author} {\bibinfo {author} {\bibfnamefont {C.}~\bibnamefont {Jing}}\ and\ \bibinfo {author} {\bibfnamefont {G.}~\bibnamefont {Ha}},\ }\bibfield  {title} {\enquote {\bibinfo {title} {Roadmap for {{Structure-based Wakefield Accelerator}} ({{SWFA}}) {{R}}\&{{D}} and its challenges in beam dynamics},}\ }\href {\doibase 10.1088/1748-0221/17/05/T05007} {\bibfield  {journal} {\bibinfo  {journal} {Journal of Instrumentation}\ }\textbf {\bibinfo {volume} {17}},\ \bibinfo {pages} {T05007} (\bibinfo {year} {2022})}\BibitemShut {NoStop}%
\bibitem [{\citenamefont {Kurz}\ \emph {et~al.}(2021)\citenamefont {Kurz}, \citenamefont {Heinemann}, \citenamefont {Gilljohann}, \citenamefont {Chang}, \citenamefont {Couperus~Cabada{\u g}}, \citenamefont {Debus}, \citenamefont {Kononenko}, \citenamefont {Pausch}, \citenamefont {Sch{\"o}bel}, \citenamefont {Assmann}, \citenamefont {Bussmann}, \citenamefont {Ding}, \citenamefont {G{\"o}tzfried}, \citenamefont {K{\"o}hler}, \citenamefont {Raj}, \citenamefont {Schindler}, \citenamefont {Steiniger}, \citenamefont {Zarini}, \citenamefont {Corde}, \citenamefont {D{\"o}pp}, \citenamefont {Hidding}, \citenamefont {Karsch}, \citenamefont {Schramm}, \citenamefont {{Martinez de la Ossa}},\ and\ \citenamefont {Irman}}]{kurz2021}%
  \BibitemOpen
  \bibfield  {author} {\bibinfo {author} {\bibfnamefont {T.}~\bibnamefont {Kurz}}, \bibinfo {author} {\bibfnamefont {T.}~\bibnamefont {Heinemann}}, \bibinfo {author} {\bibfnamefont {M.~F.}\ \bibnamefont {Gilljohann}}, \bibinfo {author} {\bibfnamefont {Y.~Y.}\ \bibnamefont {Chang}}, \bibinfo {author} {\bibfnamefont {J.~P.}\ \bibnamefont {Couperus~Cabada{\u g}}}, \bibinfo {author} {\bibfnamefont {A.}~\bibnamefont {Debus}}, \bibinfo {author} {\bibfnamefont {O.}~\bibnamefont {Kononenko}}, \bibinfo {author} {\bibfnamefont {R.}~\bibnamefont {Pausch}}, \bibinfo {author} {\bibfnamefont {S.}~\bibnamefont {Sch{\"o}bel}}, \bibinfo {author} {\bibfnamefont {R.~W.}\ \bibnamefont {Assmann}}, \bibinfo {author} {\bibfnamefont {M.}~\bibnamefont {Bussmann}}, \bibinfo {author} {\bibfnamefont {H.}~\bibnamefont {Ding}}, \bibinfo {author} {\bibfnamefont {J.}~\bibnamefont {G{\"o}tzfried}}, \bibinfo {author} {\bibfnamefont {A.}~\bibnamefont {K{\"o}hler}}, \bibinfo {author} {\bibfnamefont {G.}~\bibnamefont {Raj}}, \bibinfo {author}
  {\bibfnamefont {S.}~\bibnamefont {Schindler}}, \bibinfo {author} {\bibfnamefont {K.}~\bibnamefont {Steiniger}}, \bibinfo {author} {\bibfnamefont {O.}~\bibnamefont {Zarini}}, \bibinfo {author} {\bibfnamefont {S.}~\bibnamefont {Corde}}, \bibinfo {author} {\bibfnamefont {A.}~\bibnamefont {D{\"o}pp}}, \bibinfo {author} {\bibfnamefont {B.}~\bibnamefont {Hidding}}, \bibinfo {author} {\bibfnamefont {S.}~\bibnamefont {Karsch}}, \bibinfo {author} {\bibfnamefont {U.}~\bibnamefont {Schramm}}, \bibinfo {author} {\bibfnamefont {A.}~\bibnamefont {{Martinez de la Ossa}}}, \ and\ \bibinfo {author} {\bibfnamefont {A.}~\bibnamefont {Irman}},\ }\bibfield  {title} {\enquote {\bibinfo {title} {Demonstration of a compact plasma accelerator powered by laser-accelerated electron beams},}\ }\href {\doibase 10.1038/s41467-021-23000-7} {\bibfield  {journal} {\bibinfo  {journal} {Nature Communications}\ }\textbf {\bibinfo {volume} {12}},\ \bibinfo {pages} {2895} (\bibinfo {year} {2021})}\BibitemShut {NoStop}%
\bibitem [{\citenamefont {Foerster}\ \emph {et~al.}(2022)\citenamefont {Foerster}, \citenamefont {D{\"o}pp}, \citenamefont {Haberstroh}, \citenamefont {v.~Grafenstein}, \citenamefont {Campbell}, \citenamefont {Chang}, \citenamefont {Corde}, \citenamefont {Couperus~Cabada{\u g}}, \citenamefont {Debus}, \citenamefont {Gilljohann}, \citenamefont {Habib}, \citenamefont {Heinemann}, \citenamefont {Hidding}, \citenamefont {Irman}, \citenamefont {Irshad}, \citenamefont {Knetsch}, \citenamefont {Kononenko}, \citenamefont {{Martinez de la Ossa}}, \citenamefont {Nutter}, \citenamefont {Pausch}, \citenamefont {Schilling}, \citenamefont {Schletter}, \citenamefont {Sch{\"o}bel}, \citenamefont {Schramm}, \citenamefont {Travac}, \citenamefont {Ufer},\ and\ \citenamefont {Karsch}}]{foerster2022}%
  \BibitemOpen
  \bibfield  {author} {\bibinfo {author} {\bibfnamefont {F.~M.}\ \bibnamefont {Foerster}}, \bibinfo {author} {\bibfnamefont {A.}~\bibnamefont {D{\"o}pp}}, \bibinfo {author} {\bibfnamefont {F.}~\bibnamefont {Haberstroh}}, \bibinfo {author} {\bibfnamefont {K.}~\bibnamefont {v.~Grafenstein}}, \bibinfo {author} {\bibfnamefont {D.}~\bibnamefont {Campbell}}, \bibinfo {author} {\bibfnamefont {Y.-Y.}\ \bibnamefont {Chang}}, \bibinfo {author} {\bibfnamefont {S.}~\bibnamefont {Corde}}, \bibinfo {author} {\bibfnamefont {J.~P.}\ \bibnamefont {Couperus~Cabada{\u g}}}, \bibinfo {author} {\bibfnamefont {A.}~\bibnamefont {Debus}}, \bibinfo {author} {\bibfnamefont {M.~F.}\ \bibnamefont {Gilljohann}}, \bibinfo {author} {\bibfnamefont {A.~F.}\ \bibnamefont {Habib}}, \bibinfo {author} {\bibfnamefont {T.}~\bibnamefont {Heinemann}}, \bibinfo {author} {\bibfnamefont {B.}~\bibnamefont {Hidding}}, \bibinfo {author} {\bibfnamefont {A.}~\bibnamefont {Irman}}, \bibinfo {author} {\bibfnamefont {F.}~\bibnamefont {Irshad}}, \bibinfo
  {author} {\bibfnamefont {A.}~\bibnamefont {Knetsch}}, \bibinfo {author} {\bibfnamefont {O.}~\bibnamefont {Kononenko}}, \bibinfo {author} {\bibfnamefont {A.}~\bibnamefont {{Martinez de la Ossa}}}, \bibinfo {author} {\bibfnamefont {A.}~\bibnamefont {Nutter}}, \bibinfo {author} {\bibfnamefont {R.}~\bibnamefont {Pausch}}, \bibinfo {author} {\bibfnamefont {G.}~\bibnamefont {Schilling}}, \bibinfo {author} {\bibfnamefont {A.}~\bibnamefont {Schletter}}, \bibinfo {author} {\bibfnamefont {S.}~\bibnamefont {Sch{\"o}bel}}, \bibinfo {author} {\bibfnamefont {U.}~\bibnamefont {Schramm}}, \bibinfo {author} {\bibfnamefont {E.}~\bibnamefont {Travac}}, \bibinfo {author} {\bibfnamefont {P.}~\bibnamefont {Ufer}}, \ and\ \bibinfo {author} {\bibfnamefont {S.}~\bibnamefont {Karsch}},\ }\bibfield  {title} {\enquote {\bibinfo {title} {Stable and {{High-Quality Electron Beams}} from {{Staged Laser}} and {{Plasma Wakefield Accelerators}}},}\ }\href {\doibase 10.1103/PhysRevX.12.041016} {\bibfield  {journal} {\bibinfo  {journal}
  {Physical Review X}\ }\textbf {\bibinfo {volume} {12}},\ \bibinfo {pages} {041016} (\bibinfo {year} {2022})}\BibitemShut {NoStop}%
\bibitem [{\citenamefont {Leemans}\ \emph {et~al.}(2006)\citenamefont {Leemans}, \citenamefont {Nagler}, \citenamefont {Gonsalves}, \citenamefont {T{\'o}th}, \citenamefont {Nakamura}, \citenamefont {Geddes}, \citenamefont {Esarey}, \citenamefont {Schroeder},\ and\ \citenamefont {Hooker}}]{leemans2006}%
  \BibitemOpen
  \bibfield  {author} {\bibinfo {author} {\bibfnamefont {W.~P.}\ \bibnamefont {Leemans}}, \bibinfo {author} {\bibfnamefont {B.}~\bibnamefont {Nagler}}, \bibinfo {author} {\bibfnamefont {A.~J.}\ \bibnamefont {Gonsalves}}, \bibinfo {author} {\bibfnamefont {C.}~\bibnamefont {T{\'o}th}}, \bibinfo {author} {\bibfnamefont {K.}~\bibnamefont {Nakamura}}, \bibinfo {author} {\bibfnamefont {C.~G.~R.}\ \bibnamefont {Geddes}}, \bibinfo {author} {\bibfnamefont {E.}~\bibnamefont {Esarey}}, \bibinfo {author} {\bibfnamefont {C.~B.}\ \bibnamefont {Schroeder}}, \ and\ \bibinfo {author} {\bibfnamefont {S.~M.}\ \bibnamefont {Hooker}},\ }\bibfield  {title} {\enquote {\bibinfo {title} {{{GeV}} electron beams from a centimetre-scale accelerator},}\ }\href {\doibase 10.1038/nphys418} {\bibfield  {journal} {\bibinfo  {journal} {Nature Physics}\ }\textbf {\bibinfo {volume} {2}},\ \bibinfo {pages} {696--699} (\bibinfo {year} {2006})}\BibitemShut {NoStop}%
\bibitem [{\citenamefont {Wang}\ \emph {et~al.}(2013)\citenamefont {Wang}, \citenamefont {Zgadzaj}, \citenamefont {Fazel}, \citenamefont {Li}, \citenamefont {Yi}, \citenamefont {Zhang}, \citenamefont {Henderson}, \citenamefont {Chang}, \citenamefont {Korzekwa}, \citenamefont {Tsai}, \citenamefont {Pai}, \citenamefont {Quevedo}, \citenamefont {Dyer}, \citenamefont {Gaul}, \citenamefont {Martinez}, \citenamefont {Bernstein}, \citenamefont {Borger}, \citenamefont {Spinks}, \citenamefont {Donovan}, \citenamefont {Khudik}, \citenamefont {Shvets}, \citenamefont {Ditmire},\ and\ \citenamefont {Downer}}]{wang2013}%
  \BibitemOpen
  \bibfield  {author} {\bibinfo {author} {\bibfnamefont {X.}~\bibnamefont {Wang}}, \bibinfo {author} {\bibfnamefont {R.}~\bibnamefont {Zgadzaj}}, \bibinfo {author} {\bibfnamefont {N.}~\bibnamefont {Fazel}}, \bibinfo {author} {\bibfnamefont {Z.}~\bibnamefont {Li}}, \bibinfo {author} {\bibfnamefont {S.~A.}\ \bibnamefont {Yi}}, \bibinfo {author} {\bibfnamefont {X.}~\bibnamefont {Zhang}}, \bibinfo {author} {\bibfnamefont {W.}~\bibnamefont {Henderson}}, \bibinfo {author} {\bibfnamefont {Y.-Y.}\ \bibnamefont {Chang}}, \bibinfo {author} {\bibfnamefont {R.}~\bibnamefont {Korzekwa}}, \bibinfo {author} {\bibfnamefont {H.-E.}\ \bibnamefont {Tsai}}, \bibinfo {author} {\bibfnamefont {C.-H.}\ \bibnamefont {Pai}}, \bibinfo {author} {\bibfnamefont {H.}~\bibnamefont {Quevedo}}, \bibinfo {author} {\bibfnamefont {G.}~\bibnamefont {Dyer}}, \bibinfo {author} {\bibfnamefont {E.}~\bibnamefont {Gaul}}, \bibinfo {author} {\bibfnamefont {M.}~\bibnamefont {Martinez}}, \bibinfo {author} {\bibfnamefont {A.~C.}\ \bibnamefont {Bernstein}},
  \bibinfo {author} {\bibfnamefont {T.}~\bibnamefont {Borger}}, \bibinfo {author} {\bibfnamefont {M.}~\bibnamefont {Spinks}}, \bibinfo {author} {\bibfnamefont {M.}~\bibnamefont {Donovan}}, \bibinfo {author} {\bibfnamefont {V.}~\bibnamefont {Khudik}}, \bibinfo {author} {\bibfnamefont {G.}~\bibnamefont {Shvets}}, \bibinfo {author} {\bibfnamefont {T.}~\bibnamefont {Ditmire}}, \ and\ \bibinfo {author} {\bibfnamefont {M.~C.}\ \bibnamefont {Downer}},\ }\bibfield  {title} {\enquote {\bibinfo {title} {Quasi-monoenergetic laser-plasma acceleration of electrons to 2 {{GeV}}},}\ }\href {\doibase 10.1038/ncomms2988} {\bibfield  {journal} {\bibinfo  {journal} {Nature Communications}\ }\textbf {\bibinfo {volume} {4}},\ \bibinfo {pages} {1988} (\bibinfo {year} {2013})}\BibitemShut {NoStop}%
\bibitem [{\citenamefont {Kim}\ \emph {et~al.}(2017)\citenamefont {Kim}, \citenamefont {Pathak}, \citenamefont {Hong~Pae}, \citenamefont {Lifschitz}, \citenamefont {Sylla}, \citenamefont {Shin}, \citenamefont {Hojbota}, \citenamefont {Lee}, \citenamefont {Sung}, \citenamefont {Lee}, \citenamefont {Guillaume}, \citenamefont {Thaury}, \citenamefont {Nakajima}, \citenamefont {Vieira}, \citenamefont {Silva}, \citenamefont {Malka},\ and\ \citenamefont {Nam}}]{kim2017}%
  \BibitemOpen
  \bibfield  {author} {\bibinfo {author} {\bibfnamefont {H.~T.}\ \bibnamefont {Kim}}, \bibinfo {author} {\bibfnamefont {V.~B.}\ \bibnamefont {Pathak}}, \bibinfo {author} {\bibfnamefont {K.}~\bibnamefont {Hong~Pae}}, \bibinfo {author} {\bibfnamefont {A.}~\bibnamefont {Lifschitz}}, \bibinfo {author} {\bibfnamefont {F.}~\bibnamefont {Sylla}}, \bibinfo {author} {\bibfnamefont {J.~H.}\ \bibnamefont {Shin}}, \bibinfo {author} {\bibfnamefont {C.}~\bibnamefont {Hojbota}}, \bibinfo {author} {\bibfnamefont {S.~K.}\ \bibnamefont {Lee}}, \bibinfo {author} {\bibfnamefont {J.~H.}\ \bibnamefont {Sung}}, \bibinfo {author} {\bibfnamefont {H.~W.}\ \bibnamefont {Lee}}, \bibinfo {author} {\bibfnamefont {E.}~\bibnamefont {Guillaume}}, \bibinfo {author} {\bibfnamefont {C.}~\bibnamefont {Thaury}}, \bibinfo {author} {\bibfnamefont {K.}~\bibnamefont {Nakajima}}, \bibinfo {author} {\bibfnamefont {J.}~\bibnamefont {Vieira}}, \bibinfo {author} {\bibfnamefont {L.~O.}\ \bibnamefont {Silva}}, \bibinfo {author} {\bibfnamefont
  {V.}~\bibnamefont {Malka}}, \ and\ \bibinfo {author} {\bibfnamefont {C.~H.}\ \bibnamefont {Nam}},\ }\bibfield  {title} {\enquote {\bibinfo {title} {Stable multi-{{GeV}} electron accelerator driven by waveform-controlled {{PW}} laser pulses},}\ }\href {\doibase 10.1038/s41598-017-09267-1} {\bibfield  {journal} {\bibinfo  {journal} {Scientific Reports}\ }\textbf {\bibinfo {volume} {7}},\ \bibinfo {pages} {10203} (\bibinfo {year} {2017})}\BibitemShut {NoStop}%
\bibitem [{\citenamefont {Gonsalves}\ \emph {et~al.}(2019)\citenamefont {Gonsalves}, \citenamefont {Nakamura}, \citenamefont {Daniels}, \citenamefont {Benedetti}, \citenamefont {Pieronek}, \citenamefont {{de Raadt}}, \citenamefont {Steinke}, \citenamefont {Bin}, \citenamefont {Bulanov}, \citenamefont {{van Tilborg}}, \citenamefont {Geddes}, \citenamefont {Schroeder}, \citenamefont {T{\'o}th}, \citenamefont {Esarey}, \citenamefont {Swanson}, \citenamefont {{Fan-Chiang}}, \citenamefont {Bagdasarov}, \citenamefont {Bobrova}, \citenamefont {Gasilov}, \citenamefont {Korn}, \citenamefont {Sasorov},\ and\ \citenamefont {Leemans}}]{gonsalves2019}%
  \BibitemOpen
  \bibfield  {author} {\bibinfo {author} {\bibfnamefont {A.~J.}\ \bibnamefont {Gonsalves}}, \bibinfo {author} {\bibfnamefont {K.}~\bibnamefont {Nakamura}}, \bibinfo {author} {\bibfnamefont {J.}~\bibnamefont {Daniels}}, \bibinfo {author} {\bibfnamefont {C.}~\bibnamefont {Benedetti}}, \bibinfo {author} {\bibfnamefont {C.}~\bibnamefont {Pieronek}}, \bibinfo {author} {\bibfnamefont {T.~C.~H.}\ \bibnamefont {{de Raadt}}}, \bibinfo {author} {\bibfnamefont {S.}~\bibnamefont {Steinke}}, \bibinfo {author} {\bibfnamefont {J.~H.}\ \bibnamefont {Bin}}, \bibinfo {author} {\bibfnamefont {S.~S.}\ \bibnamefont {Bulanov}}, \bibinfo {author} {\bibfnamefont {J.}~\bibnamefont {{van Tilborg}}}, \bibinfo {author} {\bibfnamefont {C.~G.~R.}\ \bibnamefont {Geddes}}, \bibinfo {author} {\bibfnamefont {C.~B.}\ \bibnamefont {Schroeder}}, \bibinfo {author} {\bibfnamefont {{\relax Cs}.}~\bibnamefont {T{\'o}th}}, \bibinfo {author} {\bibfnamefont {E.}~\bibnamefont {Esarey}}, \bibinfo {author} {\bibfnamefont {K.}~\bibnamefont {Swanson}},
  \bibinfo {author} {\bibfnamefont {L.}~\bibnamefont {{Fan-Chiang}}}, \bibinfo {author} {\bibfnamefont {G.}~\bibnamefont {Bagdasarov}}, \bibinfo {author} {\bibfnamefont {N.}~\bibnamefont {Bobrova}}, \bibinfo {author} {\bibfnamefont {V.}~\bibnamefont {Gasilov}}, \bibinfo {author} {\bibfnamefont {G.}~\bibnamefont {Korn}}, \bibinfo {author} {\bibfnamefont {P.}~\bibnamefont {Sasorov}}, \ and\ \bibinfo {author} {\bibfnamefont {W.~P.}\ \bibnamefont {Leemans}},\ }\bibfield  {title} {\enquote {\bibinfo {title} {Petawatt {{Laser Guiding}} and {{Electron Beam Acceleration}} to 8 {{GeV}} in a {{Laser-Heated Capillary Discharge Waveguide}}},}\ }\href {\doibase 10.1103/PhysRevLett.122.084801} {\bibfield  {journal} {\bibinfo  {journal} {Physical Review Letters}\ }\textbf {\bibinfo {volume} {122}},\ \bibinfo {pages} {084801} (\bibinfo {year} {2019})}\BibitemShut {NoStop}%
\bibitem [{\citenamefont {Miao}\ \emph {et~al.}(2022)\citenamefont {Miao}, \citenamefont {Shrock}, \citenamefont {Feder}, \citenamefont {Hollinger}, \citenamefont {Morrison}, \citenamefont {Nedbailo}, \citenamefont {Picksley}, \citenamefont {Song}, \citenamefont {Wang}, \citenamefont {Rocca},\ and\ \citenamefont {Milchberg}}]{miao2022}%
  \BibitemOpen
  \bibfield  {author} {\bibinfo {author} {\bibfnamefont {B.}~\bibnamefont {Miao}}, \bibinfo {author} {\bibfnamefont {J.~E.}\ \bibnamefont {Shrock}}, \bibinfo {author} {\bibfnamefont {L.}~\bibnamefont {Feder}}, \bibinfo {author} {\bibfnamefont {R.~C.}\ \bibnamefont {Hollinger}}, \bibinfo {author} {\bibfnamefont {J.}~\bibnamefont {Morrison}}, \bibinfo {author} {\bibfnamefont {R.}~\bibnamefont {Nedbailo}}, \bibinfo {author} {\bibfnamefont {A.}~\bibnamefont {Picksley}}, \bibinfo {author} {\bibfnamefont {H.}~\bibnamefont {Song}}, \bibinfo {author} {\bibfnamefont {S.}~\bibnamefont {Wang}}, \bibinfo {author} {\bibfnamefont {J.~J.}\ \bibnamefont {Rocca}}, \ and\ \bibinfo {author} {\bibfnamefont {H.~M.}\ \bibnamefont {Milchberg}},\ }\bibfield  {title} {\enquote {\bibinfo {title} {Multi-{{GeV Electron Bunches}} from an {{All-Optical Laser Wakefield Accelerator}}},}\ }\href {\doibase 10.1103/PhysRevX.12.031038} {\bibfield  {journal} {\bibinfo  {journal} {Physical Review X}\ }\textbf {\bibinfo {volume} {12}},\ \bibinfo
  {pages} {031038} (\bibinfo {year} {2022})}\BibitemShut {NoStop}%
\bibitem [{\citenamefont {Wang}\ \emph {et~al.}(2021)\citenamefont {Wang}, \citenamefont {Feng}, \citenamefont {Ke}, \citenamefont {Yu}, \citenamefont {Xu}, \citenamefont {Qi}, \citenamefont {Chen}, \citenamefont {Qin}, \citenamefont {Zhang}, \citenamefont {Fang}, \citenamefont {Liu}, \citenamefont {Jiang}, \citenamefont {Wang}, \citenamefont {Wang}, \citenamefont {Yang}, \citenamefont {Wu}, \citenamefont {Leng}, \citenamefont {Liu}, \citenamefont {Li},\ and\ \citenamefont {Xu}}]{wang2021}%
  \BibitemOpen
  \bibfield  {author} {\bibinfo {author} {\bibfnamefont {W.}~\bibnamefont {Wang}}, \bibinfo {author} {\bibfnamefont {K.}~\bibnamefont {Feng}}, \bibinfo {author} {\bibfnamefont {L.}~\bibnamefont {Ke}}, \bibinfo {author} {\bibfnamefont {C.}~\bibnamefont {Yu}}, \bibinfo {author} {\bibfnamefont {Y.}~\bibnamefont {Xu}}, \bibinfo {author} {\bibfnamefont {R.}~\bibnamefont {Qi}}, \bibinfo {author} {\bibfnamefont {Y.}~\bibnamefont {Chen}}, \bibinfo {author} {\bibfnamefont {Z.}~\bibnamefont {Qin}}, \bibinfo {author} {\bibfnamefont {Z.}~\bibnamefont {Zhang}}, \bibinfo {author} {\bibfnamefont {M.}~\bibnamefont {Fang}}, \bibinfo {author} {\bibfnamefont {J.}~\bibnamefont {Liu}}, \bibinfo {author} {\bibfnamefont {K.}~\bibnamefont {Jiang}}, \bibinfo {author} {\bibfnamefont {H.}~\bibnamefont {Wang}}, \bibinfo {author} {\bibfnamefont {C.}~\bibnamefont {Wang}}, \bibinfo {author} {\bibfnamefont {X.}~\bibnamefont {Yang}}, \bibinfo {author} {\bibfnamefont {F.}~\bibnamefont {Wu}}, \bibinfo {author} {\bibfnamefont {Y.}~\bibnamefont
  {Leng}}, \bibinfo {author} {\bibfnamefont {J.}~\bibnamefont {Liu}}, \bibinfo {author} {\bibfnamefont {R.}~\bibnamefont {Li}}, \ and\ \bibinfo {author} {\bibfnamefont {Z.}~\bibnamefont {Xu}},\ }\bibfield  {title} {\enquote {\bibinfo {title} {Free-electron lasing at 27 nanometres based on a laser wakefield accelerator},}\ }\href {\doibase 10.1038/s41586-021-03678-x} {\bibfield  {journal} {\bibinfo  {journal} {Nature}\ }\textbf {\bibinfo {volume} {595}},\ \bibinfo {pages} {516--520} (\bibinfo {year} {2021})}\BibitemShut {NoStop}%
\bibitem [{\citenamefont {Maier}\ \emph {et~al.}(2020)\citenamefont {Maier}, \citenamefont {Delbos}, \citenamefont {Eichner}, \citenamefont {H{\"u}bner}, \citenamefont {Jalas}, \citenamefont {Jeppe}, \citenamefont {Jolly}, \citenamefont {Kirchen}, \citenamefont {Leroux}, \citenamefont {Messner}, \citenamefont {Schnepp}, \citenamefont {Trunk}, \citenamefont {Walker}, \citenamefont {Werle},\ and\ \citenamefont {Winkler}}]{maier2020}%
  \BibitemOpen
  \bibfield  {author} {\bibinfo {author} {\bibfnamefont {A.~R.}\ \bibnamefont {Maier}}, \bibinfo {author} {\bibfnamefont {N.~M.}\ \bibnamefont {Delbos}}, \bibinfo {author} {\bibfnamefont {T.}~\bibnamefont {Eichner}}, \bibinfo {author} {\bibfnamefont {L.}~\bibnamefont {H{\"u}bner}}, \bibinfo {author} {\bibfnamefont {S.}~\bibnamefont {Jalas}}, \bibinfo {author} {\bibfnamefont {L.}~\bibnamefont {Jeppe}}, \bibinfo {author} {\bibfnamefont {S.~W.}\ \bibnamefont {Jolly}}, \bibinfo {author} {\bibfnamefont {M.}~\bibnamefont {Kirchen}}, \bibinfo {author} {\bibfnamefont {V.}~\bibnamefont {Leroux}}, \bibinfo {author} {\bibfnamefont {P.}~\bibnamefont {Messner}}, \bibinfo {author} {\bibfnamefont {M.}~\bibnamefont {Schnepp}}, \bibinfo {author} {\bibfnamefont {M.}~\bibnamefont {Trunk}}, \bibinfo {author} {\bibfnamefont {P.~A.}\ \bibnamefont {Walker}}, \bibinfo {author} {\bibfnamefont {C.}~\bibnamefont {Werle}}, \ and\ \bibinfo {author} {\bibfnamefont {P.}~\bibnamefont {Winkler}},\ }\bibfield  {title} {\enquote {\bibinfo
  {title} {Decoding {{Sources}} of {{Energy Variability}} in a {{Laser-Plasma Accelerator}}},}\ }\href {\doibase 10.1103/PhysRevX.10.031039} {\bibfield  {journal} {\bibinfo  {journal} {Physical Review X}\ }\textbf {\bibinfo {volume} {10}},\ \bibinfo {pages} {031039} (\bibinfo {year} {2020})}\BibitemShut {NoStop}%
\bibitem [{\citenamefont {Gu{\'e}not}\ \emph {et~al.}(2017)\citenamefont {Gu{\'e}not}, \citenamefont {Gustas}, \citenamefont {Vernier}, \citenamefont {Beaurepaire}, \citenamefont {B{\"o}hle}, \citenamefont {Bocoum}, \citenamefont {Lozano}, \citenamefont {Jullien}, \citenamefont {{Lopez-Martens}}, \citenamefont {Lifschitz},\ and\ \citenamefont {Faure}}]{guenot2017}%
  \BibitemOpen
  \bibfield  {author} {\bibinfo {author} {\bibfnamefont {D.}~\bibnamefont {Gu{\'e}not}}, \bibinfo {author} {\bibfnamefont {D.}~\bibnamefont {Gustas}}, \bibinfo {author} {\bibfnamefont {A.}~\bibnamefont {Vernier}}, \bibinfo {author} {\bibfnamefont {B.}~\bibnamefont {Beaurepaire}}, \bibinfo {author} {\bibfnamefont {F.}~\bibnamefont {B{\"o}hle}}, \bibinfo {author} {\bibfnamefont {M.}~\bibnamefont {Bocoum}}, \bibinfo {author} {\bibfnamefont {M.}~\bibnamefont {Lozano}}, \bibinfo {author} {\bibfnamefont {A.}~\bibnamefont {Jullien}}, \bibinfo {author} {\bibfnamefont {R.}~\bibnamefont {{Lopez-Martens}}}, \bibinfo {author} {\bibfnamefont {A.}~\bibnamefont {Lifschitz}}, \ and\ \bibinfo {author} {\bibfnamefont {J.}~\bibnamefont {Faure}},\ }\bibfield  {title} {\enquote {\bibinfo {title} {Relativistic electron beams driven by {{kHz}} single-cycle light pulses},}\ }\href {\doibase 10.1038/nphoton.2017.46} {\bibfield  {journal} {\bibinfo  {journal} {Nature Photonics}\ }\textbf {\bibinfo {volume} {11}},\ \bibinfo {pages}
  {293--296} (\bibinfo {year} {2017})}\BibitemShut {NoStop}%
\bibitem [{\citenamefont {Salehi}\ \emph {et~al.}(2021)\citenamefont {Salehi}, \citenamefont {Le}, \citenamefont {Railing}, \citenamefont {Kolesik},\ and\ \citenamefont {Milchberg}}]{salehi2021}%
  \BibitemOpen
  \bibfield  {author} {\bibinfo {author} {\bibfnamefont {F.}~\bibnamefont {Salehi}}, \bibinfo {author} {\bibfnamefont {M.}~\bibnamefont {Le}}, \bibinfo {author} {\bibfnamefont {L.}~\bibnamefont {Railing}}, \bibinfo {author} {\bibfnamefont {M.}~\bibnamefont {Kolesik}}, \ and\ \bibinfo {author} {\bibfnamefont {H.~M.}\ \bibnamefont {Milchberg}},\ }\bibfield  {title} {\enquote {\bibinfo {title} {Laser-{{Accelerated}}, {{Low-Divergence}} 15-{{MeV Quasimonoenergetic Electron Bunches}} at 1 {{kHz}}},}\ }\href {\doibase 10.1103/PhysRevX.11.021055} {\bibfield  {journal} {\bibinfo  {journal} {Physical Review X}\ }\textbf {\bibinfo {volume} {11}},\ \bibinfo {pages} {021055} (\bibinfo {year} {2021})}\BibitemShut {NoStop}%
\bibitem [{\citenamefont {Goers}\ \emph {et~al.}(2015)\citenamefont {Goers}, \citenamefont {Hine}, \citenamefont {Feder}, \citenamefont {Miao}, \citenamefont {Salehi}, \citenamefont {Wahlstrand},\ and\ \citenamefont {Milchberg}}]{goers2015}%
  \BibitemOpen
  \bibfield  {author} {\bibinfo {author} {\bibfnamefont {A.~J.}\ \bibnamefont {Goers}}, \bibinfo {author} {\bibfnamefont {G.~A.}\ \bibnamefont {Hine}}, \bibinfo {author} {\bibfnamefont {L.}~\bibnamefont {Feder}}, \bibinfo {author} {\bibfnamefont {B.}~\bibnamefont {Miao}}, \bibinfo {author} {\bibfnamefont {F.}~\bibnamefont {Salehi}}, \bibinfo {author} {\bibfnamefont {J.~K.}\ \bibnamefont {Wahlstrand}}, \ and\ \bibinfo {author} {\bibfnamefont {H.~M.}\ \bibnamefont {Milchberg}},\ }\bibfield  {title} {\enquote {\bibinfo {title} {Multi-{{MeV Electron Acceleration}} by {{Subterawatt Laser Pulses}}},}\ }\href {\doibase 10.1103/PhysRevLett.115.194802} {\bibfield  {journal} {\bibinfo  {journal} {Physical Review Letters}\ }\textbf {\bibinfo {volume} {115}},\ \bibinfo {pages} {194802} (\bibinfo {year} {2015})}\BibitemShut {NoStop}%
\bibitem [{\citenamefont {Gustas}\ \emph {et~al.}(2018)\citenamefont {Gustas}, \citenamefont {Gu{\'e}not}, \citenamefont {Vernier}, \citenamefont {Dutt}, \citenamefont {B{\"o}hle}, \citenamefont {{Lopez-Martens}}, \citenamefont {Lifschitz},\ and\ \citenamefont {Faure}}]{gustas2018}%
  \BibitemOpen
  \bibfield  {author} {\bibinfo {author} {\bibfnamefont {D.}~\bibnamefont {Gustas}}, \bibinfo {author} {\bibfnamefont {D.}~\bibnamefont {Gu{\'e}not}}, \bibinfo {author} {\bibfnamefont {A.}~\bibnamefont {Vernier}}, \bibinfo {author} {\bibfnamefont {S.}~\bibnamefont {Dutt}}, \bibinfo {author} {\bibfnamefont {F.}~\bibnamefont {B{\"o}hle}}, \bibinfo {author} {\bibfnamefont {R.}~\bibnamefont {{Lopez-Martens}}}, \bibinfo {author} {\bibfnamefont {A.}~\bibnamefont {Lifschitz}}, \ and\ \bibinfo {author} {\bibfnamefont {J.}~\bibnamefont {Faure}},\ }\bibfield  {title} {\enquote {\bibinfo {title} {High-charge relativistic electron bunches from a {{kHz}} laser-plasma accelerator},}\ }\href {\doibase 10.1103/PhysRevAccelBeams.21.013401} {\bibfield  {journal} {\bibinfo  {journal} {Physical Review Accelerators and Beams}\ }\textbf {\bibinfo {volume} {21}},\ \bibinfo {pages} {013401} (\bibinfo {year} {2018})}\BibitemShut {NoStop}%
\bibitem [{\citenamefont {Salehi}\ \emph {et~al.}(2017)\citenamefont {Salehi}, \citenamefont {Goers}, \citenamefont {Hine}, \citenamefont {Feder}, \citenamefont {Kuk}, \citenamefont {Miao}, \citenamefont {Woodbury}, \citenamefont {Kim},\ and\ \citenamefont {Milchberg}}]{salehi2017}%
  \BibitemOpen
  \bibfield  {author} {\bibinfo {author} {\bibfnamefont {F.}~\bibnamefont {Salehi}}, \bibinfo {author} {\bibfnamefont {A.~J.}\ \bibnamefont {Goers}}, \bibinfo {author} {\bibfnamefont {G.~A.}\ \bibnamefont {Hine}}, \bibinfo {author} {\bibfnamefont {L.}~\bibnamefont {Feder}}, \bibinfo {author} {\bibfnamefont {D.}~\bibnamefont {Kuk}}, \bibinfo {author} {\bibfnamefont {B.}~\bibnamefont {Miao}}, \bibinfo {author} {\bibfnamefont {D.}~\bibnamefont {Woodbury}}, \bibinfo {author} {\bibfnamefont {K.~Y.}\ \bibnamefont {Kim}}, \ and\ \bibinfo {author} {\bibfnamefont {H.~M.}\ \bibnamefont {Milchberg}},\ }\bibfield  {title} {\enquote {\bibinfo {title} {{{MeV}} electron acceleration at 1\,\,{{kHz}} with {$<$}10\,\,{{mJ}} laser pulses},}\ }\href {\doibase 10.1364/OL.42.000215} {\bibfield  {journal} {\bibinfo  {journal} {Optics Letters}\ }\textbf {\bibinfo {volume} {42}},\ \bibinfo {pages} {215--218} (\bibinfo {year} {2017})}\BibitemShut {NoStop}%
\bibitem [{\citenamefont {Faure}\ \emph {et~al.}(2018)\citenamefont {Faure}, \citenamefont {Gustas}, \citenamefont {GuÃ©not}, \citenamefont {Vernier}, \citenamefont {BÃ¶hle}, \citenamefont {OuillÃ©}, \citenamefont {Haessler}, \citenamefont {Lopez-Martens},\ and\ \citenamefont {Lifschitz}}]{faure2018}%
  \BibitemOpen
  \bibfield  {author} {\bibinfo {author} {\bibfnamefont {J.}~\bibnamefont {Faure}}, \bibinfo {author} {\bibfnamefont {D.}~\bibnamefont {Gustas}}, \bibinfo {author} {\bibfnamefont {D.}~\bibnamefont {GuÃ©not}}, \bibinfo {author} {\bibfnamefont {A.}~\bibnamefont {Vernier}}, \bibinfo {author} {\bibfnamefont {F.}~\bibnamefont {BÃ¶hle}}, \bibinfo {author} {\bibfnamefont {M.}~\bibnamefont {OuillÃ©}}, \bibinfo {author} {\bibfnamefont {S.}~\bibnamefont {Haessler}}, \bibinfo {author} {\bibfnamefont {R.}~\bibnamefont {Lopez-Martens}}, \ and\ \bibinfo {author} {\bibfnamefont {A.}~\bibnamefont {Lifschitz}},\ }\bibfield  {title} {\enquote {\bibinfo {title} {A review of recent progress on laser-plasma acceleration at khz repetition rate},}\ }\href {\doibase 10.1088/1361-6587/aae047} {\bibfield  {journal} {\bibinfo  {journal} {Plasma Physics and Controlled Fusion}\ }\textbf {\bibinfo {volume} {61}},\ \bibinfo {pages} {014012} (\bibinfo {year} {2018})}\BibitemShut {NoStop}%
\bibitem [{\citenamefont {Br{\"u}mmer}\ \emph {et~al.}(2020)\citenamefont {Br{\"u}mmer}, \citenamefont {Debus}, \citenamefont {Pausch}, \citenamefont {Osterhoff},\ and\ \citenamefont {Gr{\"u}ner}}]{brummer2020}%
  \BibitemOpen
  \bibfield  {author} {\bibinfo {author} {\bibfnamefont {T.}~\bibnamefont {Br{\"u}mmer}}, \bibinfo {author} {\bibfnamefont {A.}~\bibnamefont {Debus}}, \bibinfo {author} {\bibfnamefont {R.}~\bibnamefont {Pausch}}, \bibinfo {author} {\bibfnamefont {J.}~\bibnamefont {Osterhoff}}, \ and\ \bibinfo {author} {\bibfnamefont {F.}~\bibnamefont {Gr{\"u}ner}},\ }\bibfield  {title} {\enquote {\bibinfo {title} {Design study for a compact laser-driven source for medical x-ray fluorescence imaging},}\ }\href {\doibase 10.1103/PhysRevAccelBeams.23.031601} {\bibfield  {journal} {\bibinfo  {journal} {Physical Review Accelerators and Beams}\ }\textbf {\bibinfo {volume} {23}},\ \bibinfo {pages} {031601} (\bibinfo {year} {2020})}\BibitemShut {NoStop}%
\bibitem [{\citenamefont {Svendsen}\ \emph {et~al.}(2021)\citenamefont {Svendsen}, \citenamefont {Gu{\'e}not}, \citenamefont {Svensson}, \citenamefont {Petersson}, \citenamefont {Persson},\ and\ \citenamefont {Lundh}}]{svendsen2021}%
  \BibitemOpen
  \bibfield  {author} {\bibinfo {author} {\bibfnamefont {K.}~\bibnamefont {Svendsen}}, \bibinfo {author} {\bibfnamefont {D.}~\bibnamefont {Gu{\'e}not}}, \bibinfo {author} {\bibfnamefont {J.~B.}\ \bibnamefont {Svensson}}, \bibinfo {author} {\bibfnamefont {K.}~\bibnamefont {Petersson}}, \bibinfo {author} {\bibfnamefont {A.}~\bibnamefont {Persson}}, \ and\ \bibinfo {author} {\bibfnamefont {O.}~\bibnamefont {Lundh}},\ }\bibfield  {title} {\enquote {\bibinfo {title} {A focused very high energy electron beam for fractionated stereotactic radiotherapy},}\ }\href {\doibase 10.1038/s41598-021-85451-8} {\bibfield  {journal} {\bibinfo  {journal} {Scientific Reports}\ }\textbf {\bibinfo {volume} {11}},\ \bibinfo {pages} {5844} (\bibinfo {year} {2021})}\BibitemShut {NoStop}%
\bibitem [{\citenamefont {Valenta}\ \emph {et~al.}(2020)\citenamefont {Valenta}, \citenamefont {Esirkepov}, \citenamefont {Koga}, \citenamefont {Ne{\v c}as}, \citenamefont {Grittani}, \citenamefont {Lazzarini}, \citenamefont {Klimo}, \citenamefont {Korn},\ and\ \citenamefont {Bulanov}}]{valenta2020}%
  \BibitemOpen
  \bibfield  {author} {\bibinfo {author} {\bibfnamefont {P.}~\bibnamefont {Valenta}}, \bibinfo {author} {\bibfnamefont {T.~{\relax Zh}.}\ \bibnamefont {Esirkepov}}, \bibinfo {author} {\bibfnamefont {J.~K.}\ \bibnamefont {Koga}}, \bibinfo {author} {\bibfnamefont {A.}~\bibnamefont {Ne{\v c}as}}, \bibinfo {author} {\bibfnamefont {G.~M.}\ \bibnamefont {Grittani}}, \bibinfo {author} {\bibfnamefont {C.~M.}\ \bibnamefont {Lazzarini}}, \bibinfo {author} {\bibfnamefont {O.}~\bibnamefont {Klimo}}, \bibinfo {author} {\bibfnamefont {G.}~\bibnamefont {Korn}}, \ and\ \bibinfo {author} {\bibfnamefont {S.~V.}\ \bibnamefont {Bulanov}},\ }\bibfield  {title} {\enquote {\bibinfo {title} {Polarity reversal of wakefields driven by ultrashort pulse laser},}\ }\href {\doibase 10.1103/PhysRevE.102.053216} {\bibfield  {journal} {\bibinfo  {journal} {Physical Review E}\ }\textbf {\bibinfo {volume} {102}},\ \bibinfo {pages} {053216} (\bibinfo {year} {2020})}\BibitemShut {NoStop}%
\bibitem [{\citenamefont {Huijts}\ \emph {et~al.}(2022)\citenamefont {Huijts}, \citenamefont {Rovige}, \citenamefont {Andriyash}, \citenamefont {Vernier}, \citenamefont {Ouill{\'e}}, \citenamefont {Kaur}, \citenamefont {Cheng}, \citenamefont {{Lopez-Martens}},\ and\ \citenamefont {Faure}}]{huijts2022}%
  \BibitemOpen
  \bibfield  {author} {\bibinfo {author} {\bibfnamefont {J.}~\bibnamefont {Huijts}}, \bibinfo {author} {\bibfnamefont {L.}~\bibnamefont {Rovige}}, \bibinfo {author} {\bibfnamefont {I.~A.}\ \bibnamefont {Andriyash}}, \bibinfo {author} {\bibfnamefont {A.}~\bibnamefont {Vernier}}, \bibinfo {author} {\bibfnamefont {M.}~\bibnamefont {Ouill{\'e}}}, \bibinfo {author} {\bibfnamefont {J.}~\bibnamefont {Kaur}}, \bibinfo {author} {\bibfnamefont {Z.}~\bibnamefont {Cheng}}, \bibinfo {author} {\bibfnamefont {R.}~\bibnamefont {{Lopez-Martens}}}, \ and\ \bibinfo {author} {\bibfnamefont {J.}~\bibnamefont {Faure}},\ }\bibfield  {title} {\enquote {\bibinfo {title} {Waveform {{Control}} of {{Relativistic Electron Dynamics}} in {{Laser-Plasma Acceleration}}},}\ }\href {\doibase 10.1103/PhysRevX.12.011036} {\bibfield  {journal} {\bibinfo  {journal} {Physical Review X}\ }\textbf {\bibinfo {volume} {12}},\ \bibinfo {pages} {011036} (\bibinfo {year} {2022})}\BibitemShut {NoStop}%
\bibitem [{\citenamefont {Antipenkov}\ \emph {et~al.}(2021)\citenamefont {Antipenkov}, \citenamefont {Boge}, \citenamefont {Erdman}, \citenamefont {Greco}, \citenamefont {Green}, \citenamefont {Grenfell}, \citenamefont {Hor{\'a}{\v c}ek}, \citenamefont {Hubka}, \citenamefont {Indra}, \citenamefont {Majer}, \citenamefont {Mazanec}, \citenamefont {Mazurek}, \citenamefont {Naylon}, \citenamefont {Nov{\'a}k}, \citenamefont {{\v S}obr}, \citenamefont {Spa{\v c}ek}, \citenamefont {Strkula}, \citenamefont {Szuba}, \citenamefont {Tykalewicz}, \citenamefont {Bakule},\ and\ \citenamefont {Rus}}]{antipenkov2021}%
  \BibitemOpen
  \bibfield  {author} {\bibinfo {author} {\bibfnamefont {R.}~\bibnamefont {Antipenkov}}, \bibinfo {author} {\bibfnamefont {R.}~\bibnamefont {Boge}}, \bibinfo {author} {\bibfnamefont {E.}~\bibnamefont {Erdman}}, \bibinfo {author} {\bibfnamefont {M.}~\bibnamefont {Greco}}, \bibinfo {author} {\bibfnamefont {J.~T.}\ \bibnamefont {Green}}, \bibinfo {author} {\bibfnamefont {A.}~\bibnamefont {Grenfell}}, \bibinfo {author} {\bibfnamefont {M.}~\bibnamefont {Hor{\'a}{\v c}ek}}, \bibinfo {author} {\bibfnamefont {Z.}~\bibnamefont {Hubka}}, \bibinfo {author} {\bibfnamefont {L.}~\bibnamefont {Indra}}, \bibinfo {author} {\bibfnamefont {K.}~\bibnamefont {Majer}}, \bibinfo {author} {\bibfnamefont {T.}~\bibnamefont {Mazanec}}, \bibinfo {author} {\bibfnamefont {P.}~\bibnamefont {Mazurek}}, \bibinfo {author} {\bibfnamefont {J.}~\bibnamefont {Naylon}}, \bibinfo {author} {\bibfnamefont {J.}~\bibnamefont {Nov{\'a}k}}, \bibinfo {author} {\bibfnamefont {V.}~\bibnamefont {{\v S}obr}}, \bibinfo {author} {\bibfnamefont {A.}~\bibnamefont
  {Spa{\v c}ek}}, \bibinfo {author} {\bibfnamefont {P.}~\bibnamefont {Strkula}}, \bibinfo {author} {\bibfnamefont {W.}~\bibnamefont {Szuba}}, \bibinfo {author} {\bibfnamefont {B.}~\bibnamefont {Tykalewicz}}, \bibinfo {author} {\bibfnamefont {P.}~\bibnamefont {Bakule}}, \ and\ \bibinfo {author} {\bibfnamefont {B.}~\bibnamefont {Rus}},\ }\bibfield  {title} {\enquote {\bibinfo {title} {{{TW-class Allegra Laser System}} at {{ELI-Beamlines}}},}\ }in\ \href {\doibase 10.1117/12.2592432} {\emph {\bibinfo {booktitle} {High {{Power Lasers}} and {{Applications}}}}},\ \bibinfo {editor} {edited by\ \bibinfo {editor} {\bibfnamefont {T.~J.}\ \bibnamefont {Butcher}}, \bibinfo {editor} {\bibfnamefont {J.}~\bibnamefont {Hein}}, \bibinfo {editor} {\bibfnamefont {P.}~\bibnamefont {Bakule}}, \bibinfo {editor} {\bibfnamefont {C.~L.}\ \bibnamefont {Haefner}}, \bibinfo {editor} {\bibfnamefont {G.}~\bibnamefont {Korn}}, \ and\ \bibinfo {editor} {\bibfnamefont {L.~O.}\ \bibnamefont {Silva}}}\ (\bibinfo  {publisher} {{SPIE}},\
  \bibinfo {address} {{Online Only, Czech Republic}},\ \bibinfo {year} {2021})\ p.~\bibinfo {pages} {7}\BibitemShut {NoStop}%
\bibitem [{\citenamefont {Schmid}\ \emph {et~al.}(2009)\citenamefont {Schmid}, \citenamefont {Veisz}, \citenamefont {Tavella}, \citenamefont {Benavides}, \citenamefont {Tautz}, \citenamefont {Herrmann}, \citenamefont {Buck}, \citenamefont {Hidding}, \citenamefont {Marcinkevicius}, \citenamefont {Schramm}, \citenamefont {Geissler}, \citenamefont {{Meyer-ter-Vehn}}, \citenamefont {Habs},\ and\ \citenamefont {Krausz}}]{schmid2009}%
  \BibitemOpen
  \bibfield  {author} {\bibinfo {author} {\bibfnamefont {K.}~\bibnamefont {Schmid}}, \bibinfo {author} {\bibfnamefont {L.}~\bibnamefont {Veisz}}, \bibinfo {author} {\bibfnamefont {F.}~\bibnamefont {Tavella}}, \bibinfo {author} {\bibfnamefont {S.}~\bibnamefont {Benavides}}, \bibinfo {author} {\bibfnamefont {R.}~\bibnamefont {Tautz}}, \bibinfo {author} {\bibfnamefont {D.}~\bibnamefont {Herrmann}}, \bibinfo {author} {\bibfnamefont {A.}~\bibnamefont {Buck}}, \bibinfo {author} {\bibfnamefont {B.}~\bibnamefont {Hidding}}, \bibinfo {author} {\bibfnamefont {A.}~\bibnamefont {Marcinkevicius}}, \bibinfo {author} {\bibfnamefont {U.}~\bibnamefont {Schramm}}, \bibinfo {author} {\bibfnamefont {M.}~\bibnamefont {Geissler}}, \bibinfo {author} {\bibfnamefont {J.}~\bibnamefont {{Meyer-ter-Vehn}}}, \bibinfo {author} {\bibfnamefont {D.}~\bibnamefont {Habs}}, \ and\ \bibinfo {author} {\bibfnamefont {F.}~\bibnamefont {Krausz}},\ }\bibfield  {title} {\enquote {\bibinfo {title} {Few-{{Cycle Laser-Driven Electron Acceleration}}},}\
  }\href {\doibase 10.1103/PhysRevLett.102.124801} {\bibfield  {journal} {\bibinfo  {journal} {Physical Review Letters}\ }\textbf {\bibinfo {volume} {102}},\ \bibinfo {pages} {124801} (\bibinfo {year} {2009})}\BibitemShut {NoStop}%
\bibitem [{\citenamefont {Schmid}\ \emph {et~al.}(2010)\citenamefont {Schmid}, \citenamefont {Buck}, \citenamefont {Sears}, \citenamefont {Mikhailova}, \citenamefont {Tautz}, \citenamefont {Herrmann}, \citenamefont {Geissler}, \citenamefont {Krausz},\ and\ \citenamefont {Veisz}}]{schmid2010}%
  \BibitemOpen
  \bibfield  {author} {\bibinfo {author} {\bibfnamefont {K.}~\bibnamefont {Schmid}}, \bibinfo {author} {\bibfnamefont {A.}~\bibnamefont {Buck}}, \bibinfo {author} {\bibfnamefont {C.~M.~S.}\ \bibnamefont {Sears}}, \bibinfo {author} {\bibfnamefont {J.~M.}\ \bibnamefont {Mikhailova}}, \bibinfo {author} {\bibfnamefont {R.}~\bibnamefont {Tautz}}, \bibinfo {author} {\bibfnamefont {D.}~\bibnamefont {Herrmann}}, \bibinfo {author} {\bibfnamefont {M.}~\bibnamefont {Geissler}}, \bibinfo {author} {\bibfnamefont {F.}~\bibnamefont {Krausz}}, \ and\ \bibinfo {author} {\bibfnamefont {L.}~\bibnamefont {Veisz}},\ }\bibfield  {title} {\enquote {\bibinfo {title} {Density-transition based electron injector for laser driven wakefield accelerators},}\ }\href {\doibase 10.1103/PhysRevSTAB.13.091301} {\bibfield  {journal} {\bibinfo  {journal} {Physical Review Special Topics - Accelerators and Beams}\ }\textbf {\bibinfo {volume} {13}},\ \bibinfo {pages} {091301} (\bibinfo {year} {2010})}\BibitemShut {NoStop}%
\bibitem [{\citenamefont {Lorenz}\ \emph {et~al.}(2019)\citenamefont {Lorenz}, \citenamefont {Grittani}, \citenamefont {{Chacon-Golcher}}, \citenamefont {Lazzarini}, \citenamefont {Limpouch}, \citenamefont {Nawaz}, \citenamefont {Nevrkla}, \citenamefont {Vilanova},\ and\ \citenamefont {Levato}}]{lorenz2019}%
  \BibitemOpen
  \bibfield  {author} {\bibinfo {author} {\bibfnamefont {S.}~\bibnamefont {Lorenz}}, \bibinfo {author} {\bibfnamefont {G.}~\bibnamefont {Grittani}}, \bibinfo {author} {\bibfnamefont {E.}~\bibnamefont {{Chacon-Golcher}}}, \bibinfo {author} {\bibfnamefont {C.~M.}\ \bibnamefont {Lazzarini}}, \bibinfo {author} {\bibfnamefont {J.}~\bibnamefont {Limpouch}}, \bibinfo {author} {\bibfnamefont {F.}~\bibnamefont {Nawaz}}, \bibinfo {author} {\bibfnamefont {M.}~\bibnamefont {Nevrkla}}, \bibinfo {author} {\bibfnamefont {L.}~\bibnamefont {Vilanova}}, \ and\ \bibinfo {author} {\bibfnamefont {T.}~\bibnamefont {Levato}},\ }\bibfield  {title} {\enquote {\bibinfo {title} {Characterization of supersonic and subsonic gas targets for laser wakefield electron acceleration experiments},}\ }\href {\doibase 10.1063/1.5081509} {\bibfield  {journal} {\bibinfo  {journal} {Matter and Radiation at Extremes}\ }\textbf {\bibinfo {volume} {4}},\ \bibinfo {pages} {015401} (\bibinfo {year} {2019})}\BibitemShut {NoStop}%
\bibitem [{\citenamefont {Lorenz}\ \emph {et~al.}(2020)\citenamefont {Lorenz}, \citenamefont {Grittani}, \citenamefont {Goncalves}, \citenamefont {Lazzarini}, \citenamefont {Limpouch}, \citenamefont {Nevrkla}, \citenamefont {Bulanov},\ and\ \citenamefont {Korn}}]{lorenz2020}%
  \BibitemOpen
  \bibfield  {author} {\bibinfo {author} {\bibfnamefont {S.}~\bibnamefont {Lorenz}}, \bibinfo {author} {\bibfnamefont {G.}~\bibnamefont {Grittani}}, \bibinfo {author} {\bibfnamefont {L.~V.~N.}\ \bibnamefont {Goncalves}}, \bibinfo {author} {\bibfnamefont {C.~M.}\ \bibnamefont {Lazzarini}}, \bibinfo {author} {\bibfnamefont {J.}~\bibnamefont {Limpouch}}, \bibinfo {author} {\bibfnamefont {M.}~\bibnamefont {Nevrkla}}, \bibinfo {author} {\bibfnamefont {S.}~\bibnamefont {Bulanov}}, \ and\ \bibinfo {author} {\bibfnamefont {G.}~\bibnamefont {Korn}},\ }\bibfield  {title} {\enquote {\bibinfo {title} {Tomographic reconstruction algorithms for structured gas density profiles of the targets for laser wakefield acceleration},}\ }\href {\doibase 10.1088/1361-6501/ab7cf5} {\bibfield  {journal} {\bibinfo  {journal} {Measurement Science and Technology}\ }\textbf {\bibinfo {volume} {31}},\ \bibinfo {pages} {085205} (\bibinfo {year} {2020})}\BibitemShut {NoStop}%
\bibitem [{\citenamefont {Lazzarini}\ \emph {et~al.}(2019)\citenamefont {Lazzarini}, \citenamefont {Goncalves}, \citenamefont {Grittani}, \citenamefont {Lorenz}, \citenamefont {Nevrkla}, \citenamefont {Valenta}, \citenamefont {Levato}, \citenamefont {Bulanov},\ and\ \citenamefont {Korn}}]{lazzarini2019}%
  \BibitemOpen
  \bibfield  {author} {\bibinfo {author} {\bibfnamefont {C.~M.}\ \bibnamefont {Lazzarini}}, \bibinfo {author} {\bibfnamefont {L.~V.}\ \bibnamefont {Goncalves}}, \bibinfo {author} {\bibfnamefont {G.~M.}\ \bibnamefont {Grittani}}, \bibinfo {author} {\bibfnamefont {S.}~\bibnamefont {Lorenz}}, \bibinfo {author} {\bibfnamefont {M.}~\bibnamefont {Nevrkla}}, \bibinfo {author} {\bibfnamefont {P.}~\bibnamefont {Valenta}}, \bibinfo {author} {\bibfnamefont {T.}~\bibnamefont {Levato}}, \bibinfo {author} {\bibfnamefont {S.~V.}\ \bibnamefont {Bulanov}}, \ and\ \bibinfo {author} {\bibfnamefont {G.}~\bibnamefont {Korn}},\ }\bibfield  {title} {\enquote {\bibinfo {title} {Electron acceleration at {{ELI-Beamlines}}: {{Towards}} high-energy and high-repetition rate accelerators},}\ }\href {\doibase 10.1142/S0217751X19430103} {\bibfield  {journal} {\bibinfo  {journal} {International Journal of Modern Physics A}\ }\textbf {\bibinfo {volume} {34}},\ \bibinfo {pages} {1943010} (\bibinfo {year} {2019})}\BibitemShut {NoStop}%
\bibitem [{\citenamefont {Arber}\ \emph {et~al.}(2015)\citenamefont {Arber}, \citenamefont {Bennett}, \citenamefont {Brady}, \citenamefont {{Lawrence-Douglas}}, \citenamefont {Ramsay}, \citenamefont {Sircombe}, \citenamefont {Gillies}, \citenamefont {Evans}, \citenamefont {Schmitz}, \citenamefont {Bell},\ and\ \citenamefont {Ridgers}}]{arber2015}%
  \BibitemOpen
  \bibfield  {author} {\bibinfo {author} {\bibfnamefont {T.~D.}\ \bibnamefont {Arber}}, \bibinfo {author} {\bibfnamefont {K.}~\bibnamefont {Bennett}}, \bibinfo {author} {\bibfnamefont {C.~S.}\ \bibnamefont {Brady}}, \bibinfo {author} {\bibfnamefont {A.}~\bibnamefont {{Lawrence-Douglas}}}, \bibinfo {author} {\bibfnamefont {M.~G.}\ \bibnamefont {Ramsay}}, \bibinfo {author} {\bibfnamefont {N.~J.}\ \bibnamefont {Sircombe}}, \bibinfo {author} {\bibfnamefont {P.}~\bibnamefont {Gillies}}, \bibinfo {author} {\bibfnamefont {R.~G.}\ \bibnamefont {Evans}}, \bibinfo {author} {\bibfnamefont {H.}~\bibnamefont {Schmitz}}, \bibinfo {author} {\bibfnamefont {A.~R.}\ \bibnamefont {Bell}}, \ and\ \bibinfo {author} {\bibfnamefont {C.~P.}\ \bibnamefont {Ridgers}},\ }\bibfield  {title} {\enquote {\bibinfo {title} {Contemporary particle-in-cell approach to laser-plasma modelling},}\ }\href {\doibase 10.1088/0741-3335/57/11/113001} {\bibfield  {journal} {\bibinfo  {journal} {Plasma Physics and Controlled Fusion}\ }\textbf {\bibinfo
  {volume} {57}},\ \bibinfo {pages} {113001} (\bibinfo {year} {2015})}\BibitemShut {NoStop}%
\bibitem [{\citenamefont {Bulanov}\ \emph {et~al.}(1998)\citenamefont {Bulanov}, \citenamefont {Naumova}, \citenamefont {Pegoraro},\ and\ \citenamefont {Sakai}}]{bulanov1998}%
  \BibitemOpen
  \bibfield  {author} {\bibinfo {author} {\bibfnamefont {S.}~\bibnamefont {Bulanov}}, \bibinfo {author} {\bibfnamefont {N.}~\bibnamefont {Naumova}}, \bibinfo {author} {\bibfnamefont {F.}~\bibnamefont {Pegoraro}}, \ and\ \bibinfo {author} {\bibfnamefont {J.}~\bibnamefont {Sakai}},\ }\bibfield  {title} {\enquote {\bibinfo {title} {Particle injection into the wave acceleration phase due to nonlinear wake wave breaking},}\ }\href {\doibase 10.1103/PhysRevE.58.R5257} {\bibfield  {journal} {\bibinfo  {journal} {Physical Review E}\ }\textbf {\bibinfo {volume} {58}},\ \bibinfo {pages} {R5257--R5260} (\bibinfo {year} {1998})}\BibitemShut {NoStop}%
\bibitem [{\citenamefont {Bulanov}\ \emph {et~al.}(2016)\citenamefont {Bulanov}, \citenamefont {Esirkepov}, \citenamefont {Hayashi}, \citenamefont {Kiriyama}, \citenamefont {Koga}, \citenamefont {Kotaki}, \citenamefont {Mori},\ and\ \citenamefont {Kando}}]{bulanov2016}%
  \BibitemOpen
  \bibfield  {author} {\bibinfo {author} {\bibfnamefont {S.~V.}\ \bibnamefont {Bulanov}}, \bibinfo {author} {\bibfnamefont {T.~Z.}\ \bibnamefont {Esirkepov}}, \bibinfo {author} {\bibfnamefont {Y.}~\bibnamefont {Hayashi}}, \bibinfo {author} {\bibfnamefont {H.}~\bibnamefont {Kiriyama}}, \bibinfo {author} {\bibfnamefont {J.~K.}\ \bibnamefont {Koga}}, \bibinfo {author} {\bibfnamefont {H.}~\bibnamefont {Kotaki}}, \bibinfo {author} {\bibfnamefont {M.}~\bibnamefont {Mori}}, \ and\ \bibinfo {author} {\bibfnamefont {M.}~\bibnamefont {Kando}},\ }\bibfield  {title} {\enquote {\bibinfo {title} {On some theoretical problems of laser wake-field accelerators},}\ }\href {\doibase 10.1017/S0022377816000623} {\bibfield  {journal} {\bibinfo  {journal} {Journal of Plasma Physics}\ }\textbf {\bibinfo {volume} {82}},\ \bibinfo {pages} {905820308} (\bibinfo {year} {2016})}\BibitemShut {NoStop}%
\bibitem [{\citenamefont {Hidding}\ \emph {et~al.}(2017)\citenamefont {Hidding}, \citenamefont {Karger}, \citenamefont {KÃ¶nigstein}, \citenamefont {Pretzler}, \citenamefont {Manahan}, \citenamefont {McKenna}, \citenamefont {Gray}, \citenamefont {Wilson}, \citenamefont {Wiggins}, \citenamefont {Welsh}, \citenamefont {Beaton}, \citenamefont {Delinikolas}, \citenamefont {Jaroszynski}, \citenamefont {Rosenzweig}, \citenamefont {Karmakar}, \citenamefont {Ferlet-Cavrois}, \citenamefont {Costantino}, \citenamefont {Muschitiello},\ and\ \citenamefont {Daly}}]{hidding2017}%
  \BibitemOpen
  \bibfield  {author} {\bibinfo {author} {\bibfnamefont {B.}~\bibnamefont {Hidding}}, \bibinfo {author} {\bibfnamefont {O.}~\bibnamefont {Karger}}, \bibinfo {author} {\bibfnamefont {T.}~\bibnamefont {KÃ¶nigstein}}, \bibinfo {author} {\bibfnamefont {G.}~\bibnamefont {Pretzler}}, \bibinfo {author} {\bibfnamefont {G.~G.}\ \bibnamefont {Manahan}}, \bibinfo {author} {\bibfnamefont {P.}~\bibnamefont {McKenna}}, \bibinfo {author} {\bibfnamefont {R.}~\bibnamefont {Gray}}, \bibinfo {author} {\bibfnamefont {R.}~\bibnamefont {Wilson}}, \bibinfo {author} {\bibfnamefont {S.}~\bibnamefont {Wiggins}}, \bibinfo {author} {\bibfnamefont {G.}~\bibnamefont {Welsh}}, \bibinfo {author} {\bibfnamefont {A.}~\bibnamefont {Beaton}}, \bibinfo {author} {\bibfnamefont {P.}~\bibnamefont {Delinikolas}}, \bibinfo {author} {\bibfnamefont {D.}~\bibnamefont {Jaroszynski}}, \bibinfo {author} {\bibfnamefont {J.}~\bibnamefont {Rosenzweig}}, \bibinfo {author} {\bibfnamefont {A.}~\bibnamefont {Karmakar}}, \bibinfo {author} {\bibfnamefont
  {V.}~\bibnamefont {Ferlet-Cavrois}}, \bibinfo {author} {\bibfnamefont {A.}~\bibnamefont {Costantino}}, \bibinfo {author} {\bibfnamefont {M.}~\bibnamefont {Muschitiello}}, \ and\ \bibinfo {author} {\bibfnamefont {E.}~\bibnamefont {Daly}},\ }\bibfield  {title} {\enquote {\bibinfo {title} {Laser-plasma-based space radiation reproduction in the laboratory},}\ }\href@noop {} {\bibfield  {journal} {\bibinfo  {journal} {Sci. Reports}\ }\textbf {\bibinfo {volume} {7}},\ \bibinfo {pages} {42354} (\bibinfo {year} {2017})}\BibitemShut {NoStop}%
\bibitem [{\citenamefont {Horvath}\ \emph {et~al.}(2023)\citenamefont {Horvath}, \citenamefont {Grittani}, \citenamefont {Precek}, \citenamefont {Versaci}, \citenamefont {Bulanov},\ and\ \citenamefont {Olsovcova}}]{horvath2023}%
  \BibitemOpen
  \bibfield  {author} {\bibinfo {author} {\bibfnamefont {D.}~\bibnamefont {Horvath}}, \bibinfo {author} {\bibfnamefont {G.}~\bibnamefont {Grittani}}, \bibinfo {author} {\bibfnamefont {M.}~\bibnamefont {Precek}}, \bibinfo {author} {\bibfnamefont {R.}~\bibnamefont {Versaci}}, \bibinfo {author} {\bibfnamefont {S.~V.}\ \bibnamefont {Bulanov}}, \ and\ \bibinfo {author} {\bibfnamefont {V.}~\bibnamefont {Olsovcova}},\ }\bibfield  {title} {\enquote {\bibinfo {title} {Time dynamics of the dose deposited by relativistic ultra-short electron beams},}\ }\href {\doibase 10.1088/1361-6560/ad00a3} {\bibfield  {journal} {\bibinfo  {journal} {Physics in Medicine \& Biology}\ }\textbf {\bibinfo {volume} {68}},\ \bibinfo {pages} {22NT01} (\bibinfo {year} {2023})}\BibitemShut {NoStop}%
\bibitem [{\citenamefont {Cavallone}\ \emph {et~al.}(2021)\citenamefont {Cavallone}, \citenamefont {L~Rovige}, \citenamefont {Bayart}, \citenamefont {Delorme}, \citenamefont {Vernier}, \citenamefont {Jorge}, \citenamefont {Moeckli}, \citenamefont {Deutsch}, \citenamefont {Faure},\ and\ \citenamefont {Flacco}}]{cavallone2021}%
  \BibitemOpen
  \bibfield  {author} {\bibinfo {author} {\bibfnamefont {M.}~\bibnamefont {Cavallone}}, \bibinfo {author} {\bibfnamefont {J.~H.}\ \bibnamefont {L~Rovige}}, \bibinfo {author} {\bibfnamefont {Ã.}~\bibnamefont {Bayart}}, \bibinfo {author} {\bibfnamefont {R.}~\bibnamefont {Delorme}}, \bibinfo {author} {\bibfnamefont {A.}~\bibnamefont {Vernier}}, \bibinfo {author} {\bibfnamefont {P.~G.}\ \bibnamefont {Jorge}}, \bibinfo {author} {\bibfnamefont {R.}~\bibnamefont {Moeckli}}, \bibinfo {author} {\bibfnamefont {E.}~\bibnamefont {Deutsch}}, \bibinfo {author} {\bibfnamefont {J.}~\bibnamefont {Faure}}, \ and\ \bibinfo {author} {\bibfnamefont {A.}~\bibnamefont {Flacco}},\ }\bibfield  {title} {\enquote {\bibinfo {title} {Dosimetric characterisation and application to radiation biology of a khz laser-driven electron beam},}\ }\href {\doibase 10.1007/s00340-021-07610-z} {\bibfield  {journal} {\bibinfo  {journal} {Applied Physics B}\ }\textbf {\bibinfo {volume} {127}} (\bibinfo {year} {2021}),\
  10.1007/s00340-021-07610-z}\BibitemShut {NoStop}%
\end{thebibliography}

%

\end{document}